\documentstyle[]{latexmn}\input{psfig}

\newif\ifAMStwofonts
\AMStwofontstrue

\def\G{{\cal G}}
\def\d{{\rm d}}
\def\ii{{\rm i}}
\def\ie{${\rm i.e.\ }$}
\def\etal{${\rm et\ al.\ }$}
\def\Nb{\langle N \rangle}
\def\build#1_#2^#3{\mathrel{ \mathop{\kern 0pt#1}\limits_{#2}^{#3}}}
\def\ga{\mathrel{\mathchoice {\vcenter{\offinterlineskip\halign{\hfil
$\displaystyle##$\hfil\cr>\cr\sim\cr}}}
{\vcenter{\offinterlineskip\halign{\hfil$\textstyle##$\hfil\cr>\cr\sim\cr}}}
{\vcenter{\offinterlineskip\halign{\hfil$\scriptstyle##$\hfil\cr>\cr\sim\cr}}}
{\vcenter{\offinterlineskip\halign{\hfil$\scriptscriptstyle##$\hfil
\cr>\cr\sim\cr}}}}}
\def\la{\mathrel{\mathchoice {\vcenter{\offinterlineskip\halign{\hfil
$\displaystyle##$\hfil\cr<\cr\sim\cr}}}
{\vcenter{\offinterlineskip\halign{\hfil$\textstyle##$\hfil\cr<\cr\sim\cr}}}
{\vcenter{\offinterlineskip\halign{\hfil$\scriptstyle##$\hfil\cr<\cr\sim\cr}}}
{\vcenter{\offinterlineskip\halign{\hfil$\scriptscriptstyle##$\hfil
\cr<\cr\sim\cr}}}}}  \def\phy{\phi}

\title[Extended Perturbation Theory for the Local Density Distribution 
Function]{Extended Perturbation Theory for the Local Density Distribution 
Function}
\author[Colombi et al.]{
S. Colombi$^{1,3}$, F. Bernardeau$^{2}$, F.R. Bouchet$^{3}$, and L. Hernquist$^{4,5}$\\
$^1$ CITA, 60 St George St., Toronto, ON M5S 3H8, Canada \\
$^2$ Service de Physique Th\'eorique, C.E. de Saclay,
F-91191 Gif-sur-Yvette c\'edex, France\\
$^3$ Institut d'Astrophysique de Paris, 98 bis boulevard Arago, F-75014 Paris, France\\
$^4$ Board of Studies in Astronomy and Astrophysics, University of
California, Santa~Cruz,~CA~95064,~U.S.A.\\
$^5$ Presidential Faculty Fellow}

\date{}
\pagerange{}
\pubyear{1996}

\begin{document}

\label{firstpage}

\maketitle

\begin{abstract}
Perturbation theory makes it possible to calculate the probability
distribution function (PDF) of the large scale density field in the
small variance limit, $\sigma \ll 1$.  For top hat smoothing and
scale-free Gaussian initial fluctuations, the result depends only on
the linear variance, $\sigma_{\rm linear}$, and its logarithmic
derivative with respect to the filtering scale $-(n_{\rm
linear}+3)=d\log\sigma_{\rm linear}^2/d\log \ell$ (Bernardeau 1994a).

In this paper, we measure the PDF and its low-order moments in
scale-free simulations evolved well into the nonlinear regime and
compare the results with the above predictions, assuming that the
spectral index and the variance are {\em adjustable} parameters,
$n_{\rm eff}$ and $\sigma_{\rm eff}\equiv \sigma$, where $\sigma$
is the true, nonlinear variance.  With these
additional degrees of freedom, results from perturbation theory
provide a good fit of the PDFs, even in the highly nonlinear regime.
The value of $n_{\rm eff}$ is of course equal to $n_{\rm linear}$ when
$\sigma \ll 1$, and it decreases with increasing $\sigma$. A nearly
flat plateau is reached when $\sigma \gg 1$. In this regime, the
difference between $n_{\rm eff}$ and $n_{\rm linear}$ increases when
$n_{\rm linear}$ decreases.  For initial power-spectra with $n_{\rm
linear}=-2,-1,0,+1$, we find $n_{\rm eff} \simeq -9,-3,-1,-0.5$ when
$\sigma^2 \simeq 100$.

It is worth noting that $-(3+n_{\rm eff})$ is {\em different} from the
logarithmic derivative of the nonlinear variance with respect to the
filtering scale.  Consequently, it is not straightforward to determine
the nonlinearly evolved PDF from arbitrary (scale-dependent) initial
conditions, such as Cold Dark Matter, although we propose a simple
method that makes this feasible.

Thus, estimates of the variance (using, for example, the prescription
proposed by Hamilton et al. 1991) and of $n_{\rm eff}$ as functions of
scale for a given power spectrum makes it possible to calculate the
local density PDF at any time from the initial conditions.

\end{abstract}


\section{Introduction}

It is generally believed that the large-scale structures of the
Universe, such as those seen in the distribution of galaxies, arose
through gravitational instability from primordial seed fluctuations.
On scales larger than roughly 8 $h^{-1}$Mpc\footnote{We adopt
$h=H_0/(100\ {\rm km/s/Mpc})$, where $H_0$ is the Hubble constant.},
the dynamics of the Universe is dominated by gravitational forces,
implying that the matter behaves as if it were purely nonbaryonic and
collision-less.  Thus, when curvature can be neglected (for scales
below 500$h^{-1}$Mpc), the evolution of the Universe can be described
by the Vlasov-Poisson system of equations, in coordinates comoving
with the cosmic expansion.  In general, nonlinear coupling makes it
virtually impossible to solve these equations analytically.  Thus, on
small scales, where density contrasts are large, $N$-body simulations
are typically used to obtain numerical solutions.  However, on scales
larger than about 10$h^{-1}$Mpc, the amplitude of the density
fluctuations is modest and {\em Perturbation Theory} (PT) can be used
to compute statistical properties of the cosmic fields.

Here, we focus on the Probability Distribution Function (PDF) of the
local density smoothed with a spherical filter of radius $\ell$ at a
time $t$, $P(\rho,\ell,t)$.  The PDF, a simple to measure statistic, 
has been widely used to characterize
the statistical properties of the
large scale galaxy distribution (see, e.g., Alimi, Blanchard \&
Schaeffer 1990; Maurogordato, Schaeffer \& da Costa 1992;
Szapudi, Szalay \& Bosch\'an 1992;
Bouchet et al. 1993; Gazta\~naga 1992, 1994; Szapudi, Meiksin \&
Nichol 1996). 

Our aim is to determine whether there is a
simple relationship between the PDF obtained for the
matter distribution in the nonlinear regime
and the one predicted by PT. The latter should, in principle, be valid
only in the weakly nonlinear regime.  In our calculations, we assume
i) that the Universe is flat with $\Omega = 1$, and ii) that the
initial fluctuations of the density field were Gaussian, pressure-less
(i.e. cold), and scale-free. 

In real galaxy catalogs, the interpretation of the
results is complicated by systematic effects,
such as the so-called bias between the galaxy
distribution and the matter distribution (e.g., Fry \& Gazta\~naga
1993, Juszkiewicz et al. 1995), redshift distortions in
three-dimensional catalogs (e.g., Matsubara \& Suto 1994; Hivon et
al. 1995), and effects of projection along the line of sight in
two-dimensional catalogs (e.g., Bernardeau 1995). In what follows,
we focus on the statistical properties of the {\em matter}
distribution and thus neglect these effects. 

Let $\sigma^2(\ell)$ be the variance of the PDF.  In the weakly
nonlinear regime ($\sigma^2 \ll 1$) the scale-free nature of the
initial conditions implies
\begin{equation}
  \sigma^2\approx\sigma^2_{\rm linear} \propto \ell^{-(n_{\rm linear}+3)}.
\end{equation}
In general, the PDF of the local density can be characterized by the
behavior of its cumulants (\ie the connected part of the moments),
$\langle \delta^q \rangle_{\rm c}$ (see e.g. Bernardeau 1994a,
hereafter B94 for a more precise definition). 
Note that $\langle
\delta^2 \rangle_{\rm c}= \langle \delta^2 \rangle=\sigma^2(\ell)$,
where $\delta$ is the density contrast $\delta = \rho/{\langle \rho
\rangle}-1$.  
 When the variance is small, the dominant contribution to
all the cumulants can be calculated analytically from PT for a
top-hat smoothing window (B94). The result is that the quantities
$S_q$ defined by
\begin{equation}
  S_q(\sigma)
\equiv \frac{\langle \delta^q \rangle_{\rm c} }{\sigma^{2(q-1)}},
 \label{eq:sqdef}
\end{equation}
are all {\em finite constants} in the 
limit of small variance\footnote{Note that we have exploited the 
scale-free nature of the system to express the quantity $S_q(\ell,t)$ as a
function of only the variance. This would not be possible for 
scale dependent initial conditions such as Cold Dark Matter (CDM).}.
The next-to-leading order term is expected to be proportional 
to $\sigma^2$, so that
\begin{equation}
   S_q(\sigma)=S_q^{\rm PT}+{\cal{O}}(\sigma^2).
   \label{eq:Sqpt}
\end{equation}
Comparisons with numerical simulations have shown that the PT
predictions are accurate as long as $\sigma<1$ (Juszkiewicz et al.
1993, 1995; B94; Baugh, Gazta\~naga \& Efstathiou 1995; Gazta\~naga \&
Baugh 1995; Lokas et al. 1995a,b).  In this regime, the whole PDF can
then be calculated analytically (B94), as a function of $\ell$,
$\sigma_{\rm linear}$ and $n_{\rm linear}$:
\begin{equation}
  P=P_{\rm PT}(\rho,\sigma_{\rm linear},n_{\rm linear}),\quad \sigma \ll 1.
\end{equation}

It is important to notice that 
the measured cumulants in the observed galaxy distribution
are also in good agreement with PT predictions in
the weakly nonlinear regime (e.g., Bouchet et al. 1993, 
Gazta\~naga 1994, Szapudi, Meiksin \&
Nichol 1996), given possible 
statistical uncertainties (e.g., Szapudi \& Colombi 1996)
and other systematic effects discussed earlier in this
section. 

For power-law spectra, the set of $S_q^{\rm PT}$ can be determined as
a function of $n_{\rm linear}$ (Juszkiewicz \& Bouchet 1992;
Juszkiewicz, Bouchet \& Colombi 1993; B94; Bernardeau 1994b):
\begin{equation}
   S_3^{\rm PT}(n_{\rm linear})=\frac{34}{7}-(n_{\rm linear}+3); \label{eq:S3per}
\end{equation}
\begin{eqnarray}
   S_4^{\rm PT}(n_{\rm linear})&=&\frac{60712}{1323}-
\frac{62}{3}(n_{\rm linear}+3) \nonumber\\
   &+&\frac{7}{3} (n_{\rm linear}+3)^2;
   \label{eq:S4per}
\end{eqnarray}
\begin{eqnarray}
   S_5^{\rm PT}(n_{\rm linear})&= &\frac{200575880}{305613}-
   \frac{1847200}{3969}(n_{\rm linear}+3)\nonumber\\
   &+&\frac{6940}{63}(n_{\rm linear}+3)^2 \nonumber\\
   &-&\frac{235}{27}(n_{\rm linear}+3)^3,  
\label{eq:S5per}
\end{eqnarray}
and so on. 

In what follows, we use these definitions as fitting functions of the
$S_q$ ratios, leaving $n_{\rm linear}$ as a free parameter. One can
then invert the above equations to deduce, for each $q$, the value
of $n_{\rm linear}$ which yields the measured value, $S_q^{\rm meas}$,
of the parameters $S_q$.  We use this approach to define functions
$n_q(\sigma)$, ($q \geq 3$), by
\begin{equation} 
   S_q^{\rm PT}[n_q(\sigma)]\equiv S_q^{\rm meas}(\sigma).  
\end{equation}
Of course, as discussed above, the measured $n_q$'s are expected
to have identical values in the weakly nonlinear regime ($\sigma \ll
1$), and be equal to the value of the initial index $n_{\rm linear}$.

In the nonlinear regime ($\sigma \ga 1$), however, results from
perturbation theory are no longer valid, and we do not expect the
functions $n_q$ to take a universal (constant) value.  Actually, there
is no {\em a priori} reason why the relation $n_q(\sigma) =
n_{q'}(\sigma)$ should hold for $q \neq q'$ and $\sigma \ga 1$, even
in the framework of the self-similar/stable clustering hypothesis,
where the functions $n_q(\sigma)$ are expected to approach constants
in the highly nonlinear regime (e.g., Davis \& Peebles 1977; Peebles
1980; Balian \& Schaeffer 1989, hereafter BS).

In this paper, we use measurements of the PDF by Colombi, Bouchet \&
Hernquist (1996, hereafter CBH) in $N$-body simulations of flat
universes with scale-free initial conditions to show that, within the
accuracy of the calculations, all functions $n_q$ superpose to give a
unique function $n_{\rm eff}$
\begin{equation}
   n_{\rm eff}(\sigma) = n_q(\sigma). 
   \label{eq:neff}
\end{equation}
The measurement of $S_q$ is difficult when $q$ is larger than a
few, because of systematic errors due to finite
volume effects (e.g., Colombi, Bouchet \& Schaeffer 1994, 1995,
hereafter CBSI and CBSII).  However, one can check the validity of the
property (\ref{eq:neff}) from the overall {\em shape} of the PDF
rather than with its moments. This would mean that PT can be applied
to the nonlinear regime as follows:
\begin{equation}
   P=P_{\rm PT}[\rho,\sigma,n_{\rm eff}(\sigma)].
   \label{eq:ept}
\end{equation}
We shall refer to this generalization as {\em Extended Perturbation
Theory} (EPT).  The only quantities that would be needed to compute
the PDF, if EPT applies, are the functions $n_{\rm eff}(\sigma)$ and
the variance itself. If valid, EPT would thus result in tremendous
simplification.

This paper is organized as follows.  In Section 2, we present the
measured quantities $n_q$, $3 \leq q \leq 5$ as functions of the
variance in our $N$-body simulations and show that they indeed define
a unique function $n_{\rm eff}(\sigma)$ that can be fitted
analytically.  We then compare the PDF predicted by EPT to that
measured for different values of $\sigma$.  The results allow an
extrapolation to non power-law spectra.  In section 3 we discuss the
asymptotic behavior of the density PDFs in the EPT framework which
depends only on a few parameters. We relate them to the $n_{\rm eff}$
found. We conclude in section 4 by summarizing our results and discuss
them in light of related works.  For completeness, in the appendix we
summarize the results obtained from PT, the generic properties
expected for the density PDF, and the algorithm used to compute it
numerically.

\section{Measurements}
The scale-free simulations we analyze are described in detail in CBH.
They were performed with the cosmological tree-code of Hernquist,
Bouchet \& Suto (1991), using $64^3$ particles and periodic
boundaries.  We consider four values of the spectral index: $n_{\rm
linear}=-2$, $-1$ (two $N$-body sets), 0 and +1.  CBH measured the
count probability distribution function (CPDF) $P_N(\ell)$ at various
times in these simulations.  The CPDF is just the probability of
finding $N$ particles in a spherical cell of radius $\ell$ randomly
thrown in the data set.  Thus, the CPDF differs from the PDF:
discreteness effects have to be accounted for to transform the CPDF
into the PDF since the latter involves the underlying {\em continuous}
density field.  CBH computed the quantities $S_q$, $3 \leq q \leq 5$
from the moments of the CPDF, with the proper corrections for
discreteness effects.  They studied other contamination effects, such
as artificial correlations related to the way initial conditions are
generated, short range softening of the forces in the $N$-body
simulations, and finite volume effects arising from missing waves
longer than the box.  In particular, CBH defined a reliable scale
range over which finite volume corrections were applied.  From these
measurements of the functions $S_q(\sigma)$, we can directly infer
$n_q(\sigma)$ for $3 \leq q \leq 5$.

This section is organized as follows. First we study the functions
$n_q(\sigma)$ in \S~2.1. We will see that they all superpose to a
single function $n_{\rm eff}(\sigma)$ within the measurement errors.
In \S~2.2, we check whether this result applies to the CPDF itself, and
in \S~2.3, we present a possible extrapolation to non power-law
spectra.

\subsection{The effective spectral index $n_{\rm eff}$} 
%

%
%
\begin{figure*}
\centerline{
\hbox{
\psfig{figure=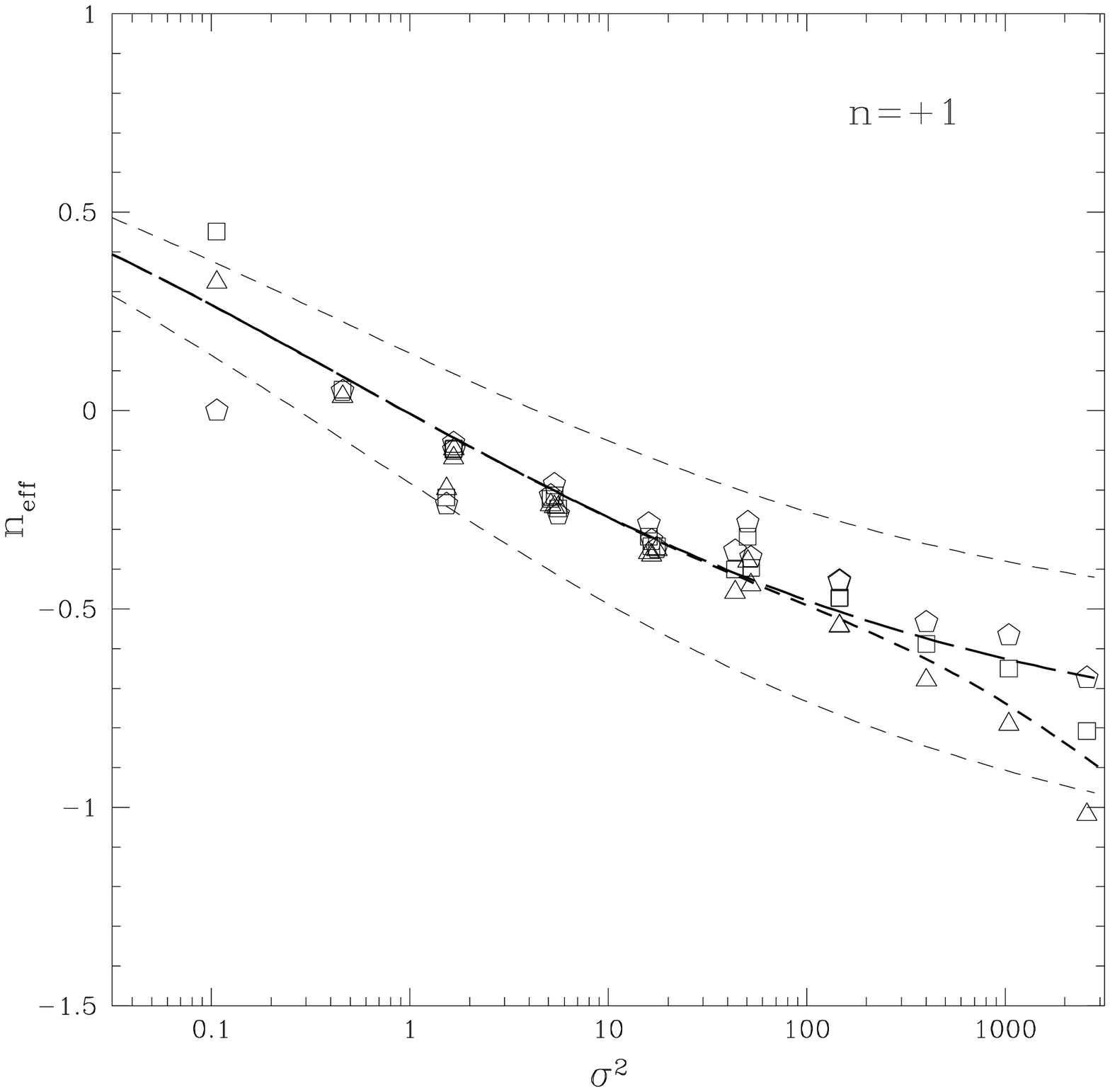,width=8.5cm}
\psfig{figure=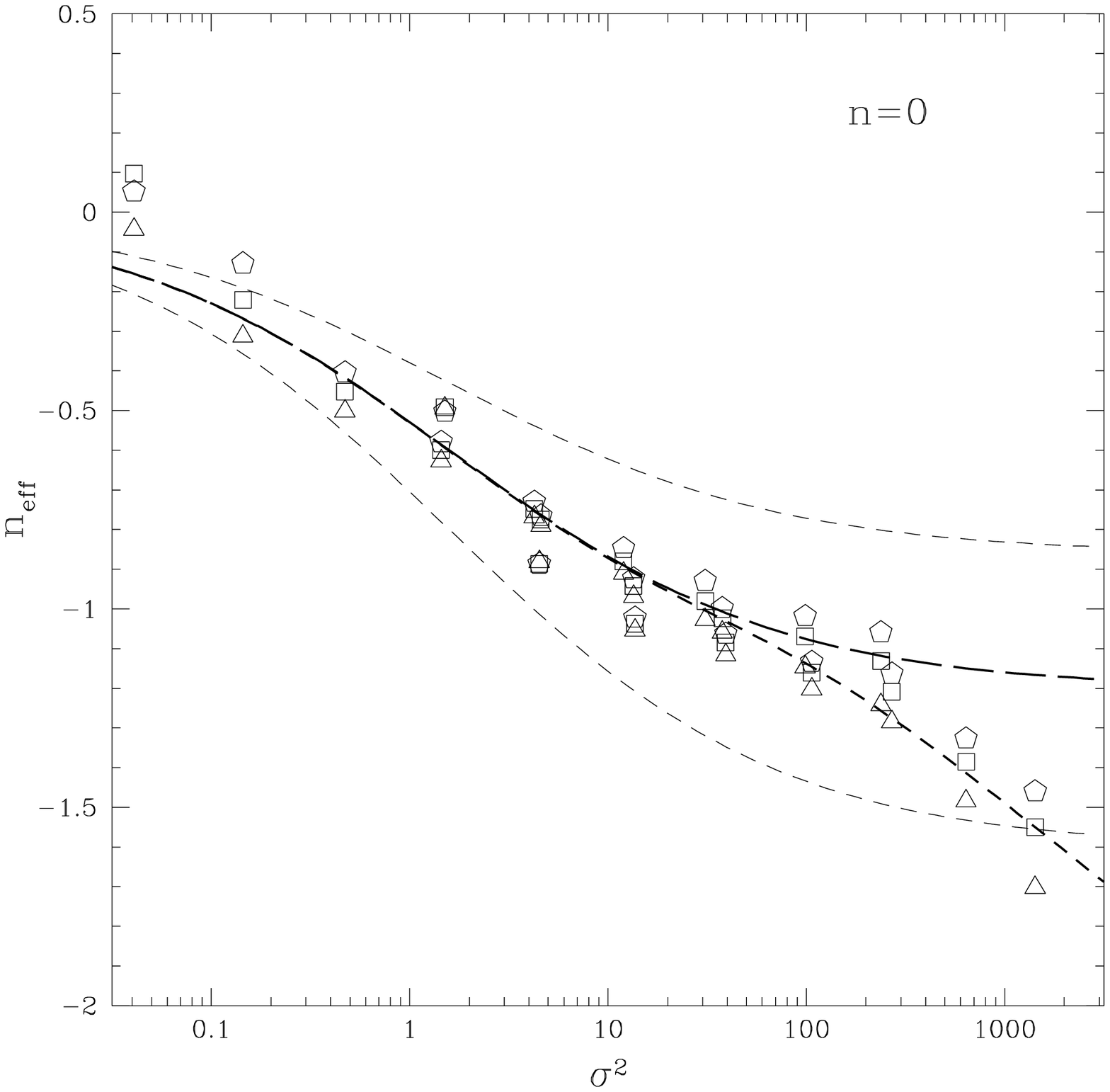,width=8.5cm}}}
\centerline{\hbox{
\psfig{figure=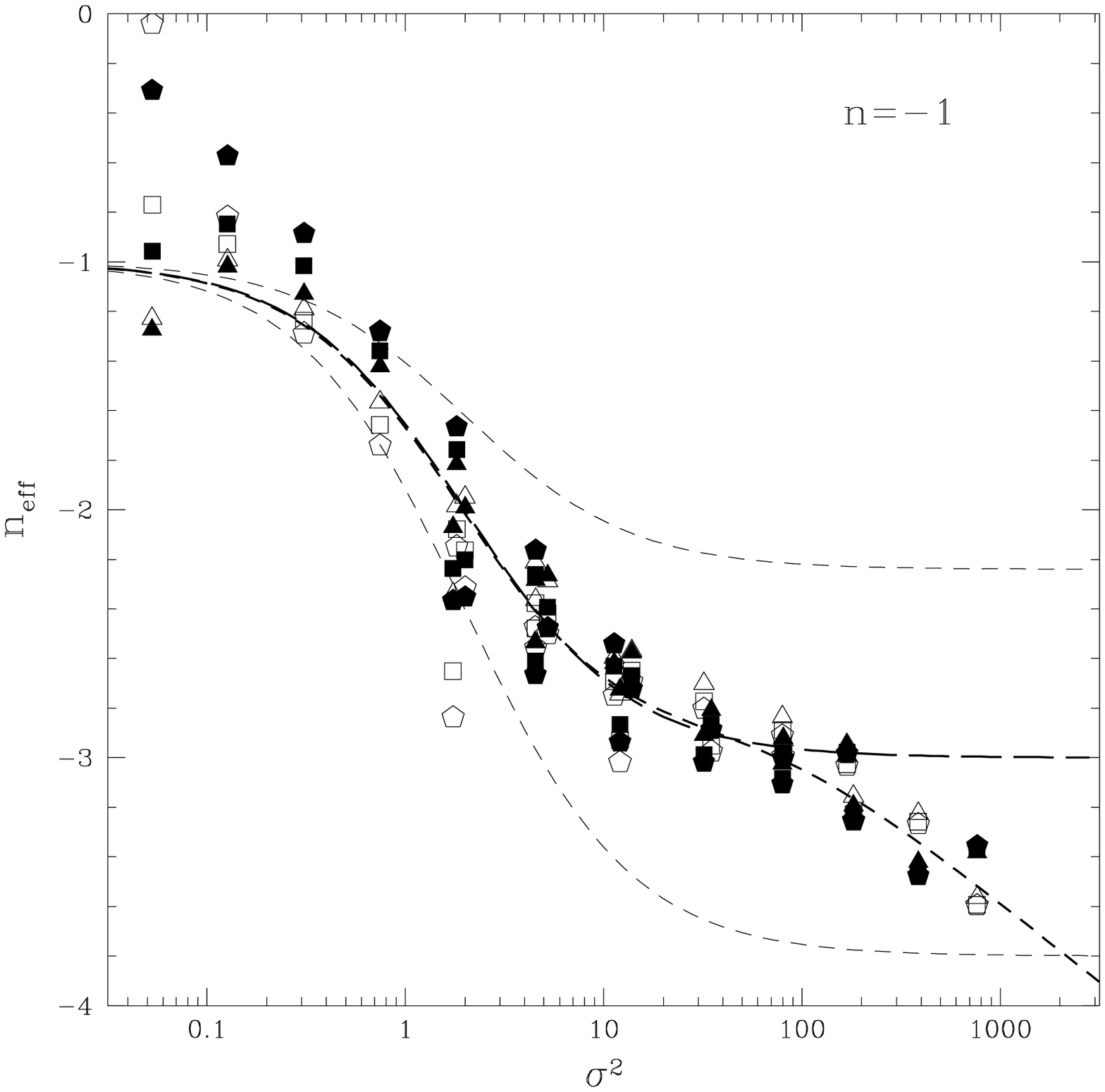,width=8.5cm}
\psfig{figure=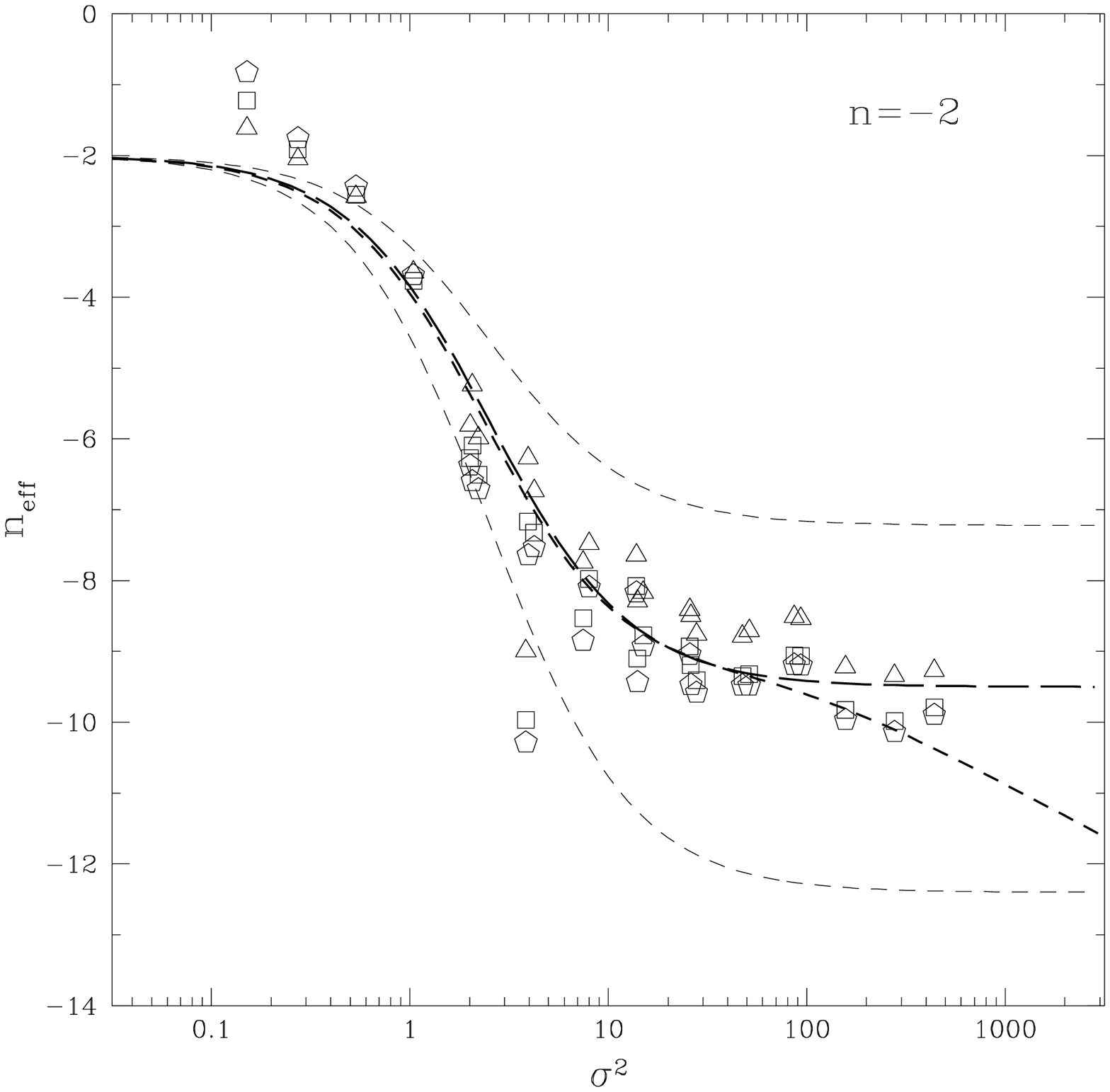,width=8.5cm}}}
\caption[ ]{The measured spectral indices $n_q$ obtained by
inverting equations (\ref{eq:S3per})
(triangles), (\ref{eq:S4per}) (squares) and 
(\ref{eq:S5per}) (pentagons), using the values of $S_q^{\rm meas}$ measured
by CBH. Each panel corresponds to a different value of the initial spectral 
index $n_{\rm linear}$. In the case $n_{\rm linear}=-1$, two
simulations have been used (open and filled symbols).
In each panel, the thick long dashed curve uses
equation (\ref{eq:fitber}) with the parameters given in 
Table~\ref{table:table1}. The thin short-dashed curves bracket the range of 
possible fits, given the uncertainties in the measurements (see text). 
The thick short-dashed curves use the prejudice (\ref{eq:fitcol})
for the large $\sigma$ behavior of $n_{\rm nonlinear}$ (see text).}
\label{fig:figure1} 
\end{figure*} 

\noindent 
In Figure \ref{fig:figure1}, we show the dependence of the deduced
$n_q$ on the variance, for each value of $n_{\rm linear}$.  The
various symbols correspond to different values of $q$. They superpose,
which is not surprising in the weakly nonlinear regime since all
functions $n_q$ should tend to the same value $n_{\rm linear}$
according to PT (although not perfectly; this point is discussed in
detail by CBH).  More interestingly, the superposition still holds in
the nonlinear regime where the vertical spread of the points {\em does
not increase} (at least not significantly), which corresponds to a
very peculiar behavior of the statistics in the nonlinear regime, as
discussed in the introduction.

In figure \ref{fig:figure1}, some analytical fits based 
on two different approaches are also shown. 
In each panel, the thick long dashes were calculated from
\begin{eqnarray}
  n_{\rm eff}^{\rm I}(\sigma)&=& n_{\rm linear} + \nonumber \\
  & & (n_{\rm nonlinear}-
n_{\rm linear}) x^\tau/(x^\tau+x^{-\tau}),
  \label{eq:fitber}
\end{eqnarray}
with
\begin{equation}
  x=\exp \left[ \log_{10} (\sigma^2/\sigma_0^2) \right].
  \label{eq:xsigm}
\end{equation}
Table \ref{table:table1} gives the corresponding values of the parameters 
$n_{\rm non linear}$, $\tau$ and $\sigma_0$
as functions of $n_{\rm linear}$. 
In equation (\ref{eq:fitber}), we impose that the index $n_{\rm eff}$ reaches
a constant plateau at large $\sigma$, $n_{\rm nonlinear}$. 
The parameter $\sigma_0$ gives the position of the transition
between the linear value and the nonlinear value, and $\tau$ defines
the width of this transition.

\begin{table}
\caption[ ]{Parameters used in fit I (eq.~[\ref{eq:fitber}]).}
\label{table:table1}
\begin{tabular}{cccccc}
\hline
$n_{\rm linear}$ & 
$n_{\rm nonlinear}$ & $n_{\rm nonlinear}^-$ & 
$n_{\rm nonlinear}^+$ & $\sigma_0$ & $\tau$ \\ \hline
   -2            &  -9.5  & -12.4 & -7.22 &  1.6  & 1.4       \\
   -1            &  -3    & -3.8  & -2.24 &  1.4  & 1.2       \\
    0            &  -1.2  & -1.6  & -0.86 &  1.25 & 0.6       \\
   +1            &  -0.85 & -1.17 & -0.57 &  0.7  & 0.3       \\ \hline
\end{tabular}
\end{table}

Alternatively it is possible to allow $n_{\rm nonlinear}$ to have a
residual $\sigma$ dependence like the one found in CBH.  The thick
short dashes in Figure~\ref{fig:figure1} use the following
interpolation formula
\begin{eqnarray}
   n_{\rm eff}^{\rm II}(\sigma) &=& n_{\rm eff}^{\rm I}(\sigma) + 
\nonumber \\ & & \left[ n_{\rm nonlinear}(\sigma)-
   n_{\rm eff}^{\rm I}(\sigma) \right] x/(x+1/x).
   \label{eq:fitcol1}
\end{eqnarray}
The parameter $x$ is given by equation (\ref{eq:xsigm}) with
$\log_{10} \sigma_0^2 =1.8$, $1.8$, $2.2$ and $3$
respectively for $n=-2$, $-1$, 0 and $+1$, and the function $n_{\rm nonlinear}(\sigma)$ is
\begin{equation}
   n_{\rm nonlinear}(\sigma)={34\over7}-3-{\widetilde S_3} (\sigma^2/100)^{0.045},
   \label{eq:fitcol}
\end{equation}
where ${\widetilde S}_3$ is the value of $S_3^{\rm meas}$ for
$\sigma=10$ found by CBH.  (We have ${\widetilde S}_3 = 10.3$, $4.8$,
$3$ and $2.4$ respectively for $n_{\rm linear}=-2$, $-1$, 0 and $+1$).

The two fits (\ref{eq:fitber}) and (\ref{eq:fitcol1}), designated
hereafter respectively by I and II, are thus based on different
theoretical prejudices.  They both suppose that PT predictions are
verified in the weakly nonlinear regime $\sigma^2 \ll 1$.  Model I
assumes moreover that the stable clustering hypothesis applies for
$\sigma^2 \gg 1$, i.e.~that the functions $S_q$ do not depend on scale
(or on $\sigma$) in the highly nonlinear regime. Of course, this
implies that $n_{\rm eff}$ becomes asymptotically constant at large
$\sigma$, and equal to $n_{\rm nonlinear}$.  Model II takes into account
the slight deviation from a constant found by CBH for the functions
$S_q^{\rm meas}$ in the highly nonlinear regime, namely
\begin{equation}
   S_q^{\rm meas} \propto \sigma^{2 \times 0.045(q-2)}.
\end{equation}
Note that such a power-law behavior is actually incompatible with the
existence of a universal function $n_{\rm eff}(\sigma)$. On the other
hand, this inconsistency is very mild and produces deviations from
property (\ref{eq:neff}) well below the uncertainties in the
measurements. Such disagreement could explain, however, the slight
broadening of the vertical scatter of the points at very large
$\sigma$ for $n=0$ and $n=+1$.

Figure \ref{fig:figure1} also contains two thin short-dashed curves
above and below the thick long-dashed one. They use the same
parameterization (\ref{eq:fitber}), but with different values of
$n_{\rm nonlinear}$, which we denote $n_{\rm nonlinear}^{-}$ and
$n_{\rm nonlinear}^{+}$ for the upper and the lower curves
respectively. To some extent, they bracket the domain of possible
values of $n_{\rm eff}$, given the uncertainties in the measurements.
The numbers $n_{\rm nonlinear}^{\pm}$ come from estimates of the
errors in the measurements $S_q^{\rm meas}$ of the parameters $S_q$ by
CBH as follows. In the nonlinear regime, such errors can be described
as $S_q^{\rm meas}(\sigma)=S_q(\sigma) f_q^{\pm 1} \equiv S_q^{\pm}$.
The values of the uncertainty factors $f_q$ are given by CBH. We take
$n_{\rm nonlinear}^{\pm} \equiv \langle n_q ^{\pm} \rangle_{q=3,4,5}$
where $n_q^{\pm}$ is the function inferred from $S_q^{\pm}$, assuming
that $S_q$ is calculated from equations (\ref{eq:S3per}),
(\ref{eq:S4per}) and (\ref{eq:S5per}) by replacing $n_{\rm linear}$
with the value of $n_{\rm nonlinear}$ given in Table
\ref{table:table1}.

The lower panels of Figure \ref{fig:figure1} contain symbols that are
significantly outside the range defined by the two thin dashed curves,
at $\sigma^2 \ll 1$ and for $\sigma^2$ of the order of few units. In
the first case, this is because our estimate of the errors tends to
zero in the weakly nonlinear regime, since we assume that $n_{\rm
eff}$ reaches $n_{\rm linear}$.  In the second case, the method used
by CBH to correct for finite volume effects can be used only when
$\sigma^2$ is much larger than unity. As a result, it is quite
difficult to apply it accurately in the regime where $\sigma^2$ is
only a few units.

\subsection{The probability distribution function} 
%
%
\begin{figure*}
\centerline{
\hbox{
\psfig{figure=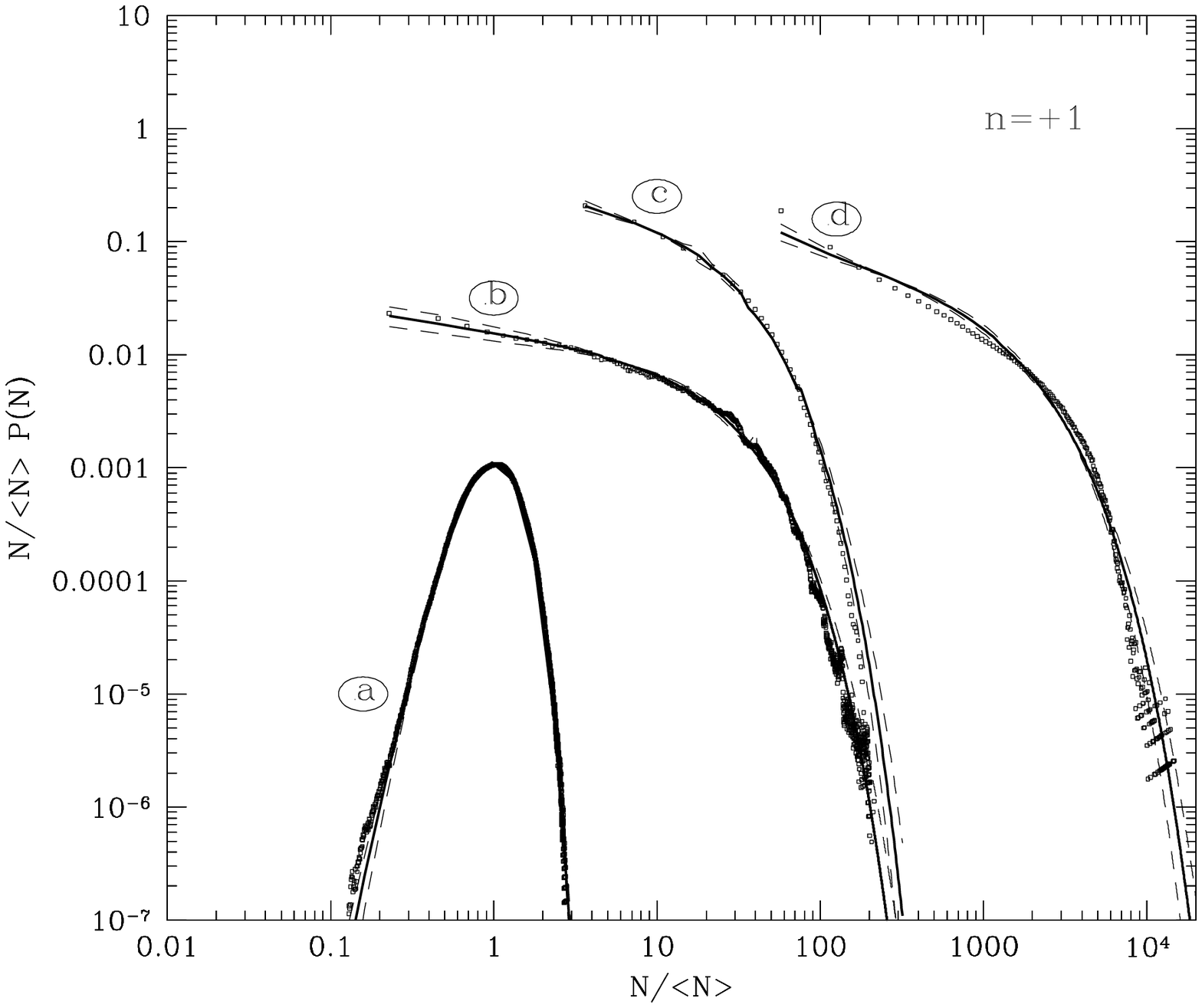,width=8.5cm}
\psfig{figure=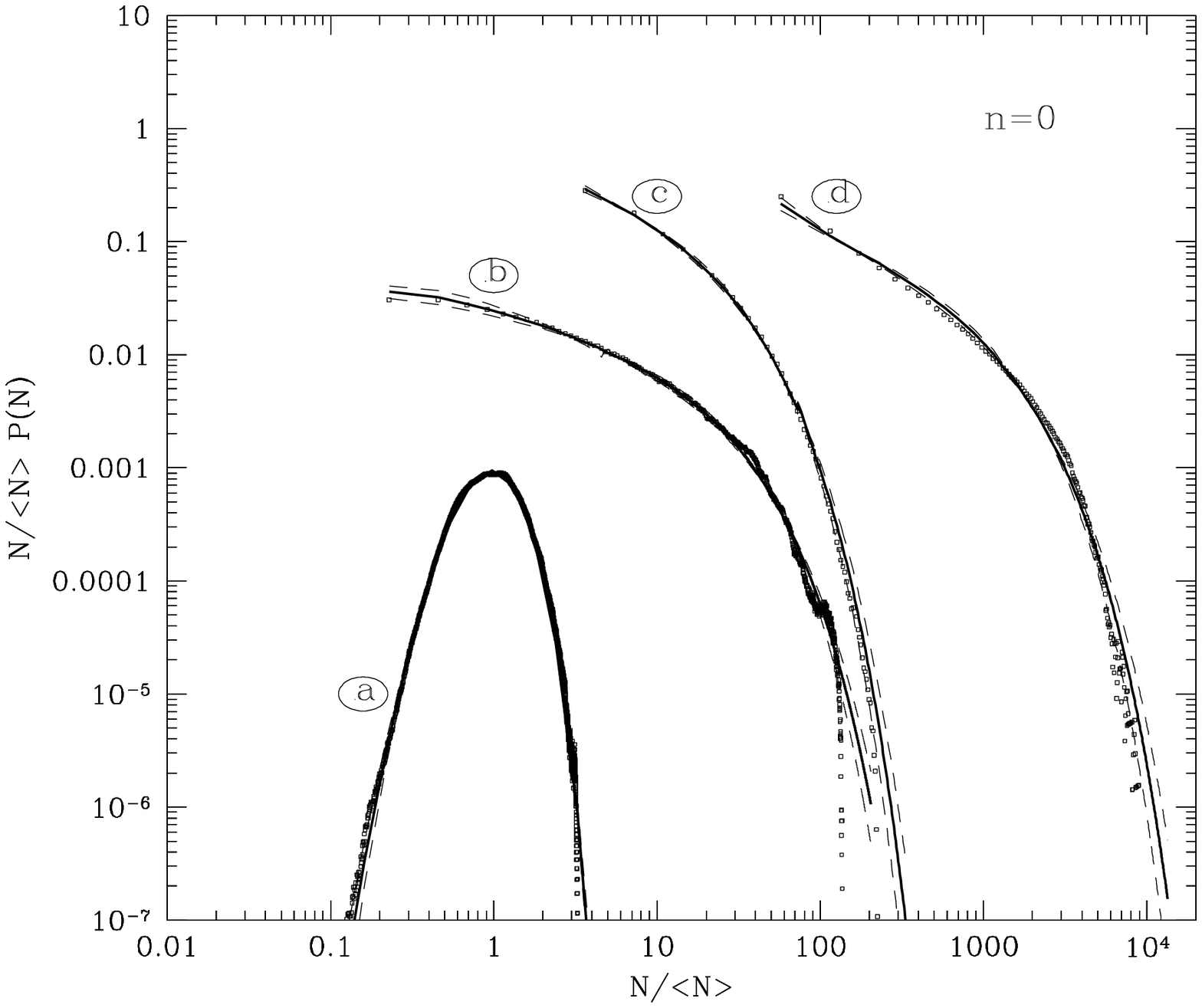,width=8.5cm}
}}
\centerline{
\hbox{
\psfig{figure=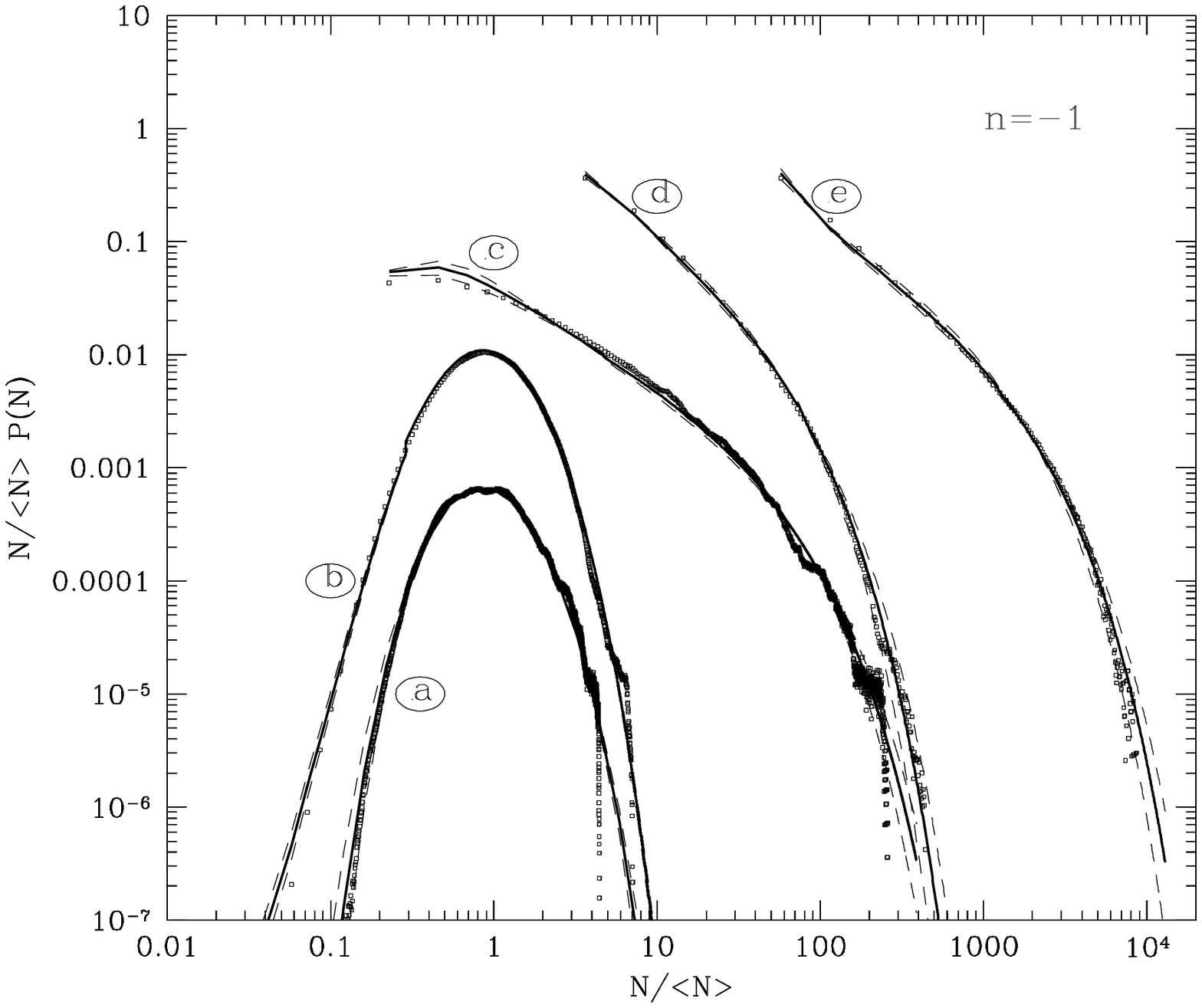,width=8.5cm}
\psfig{figure=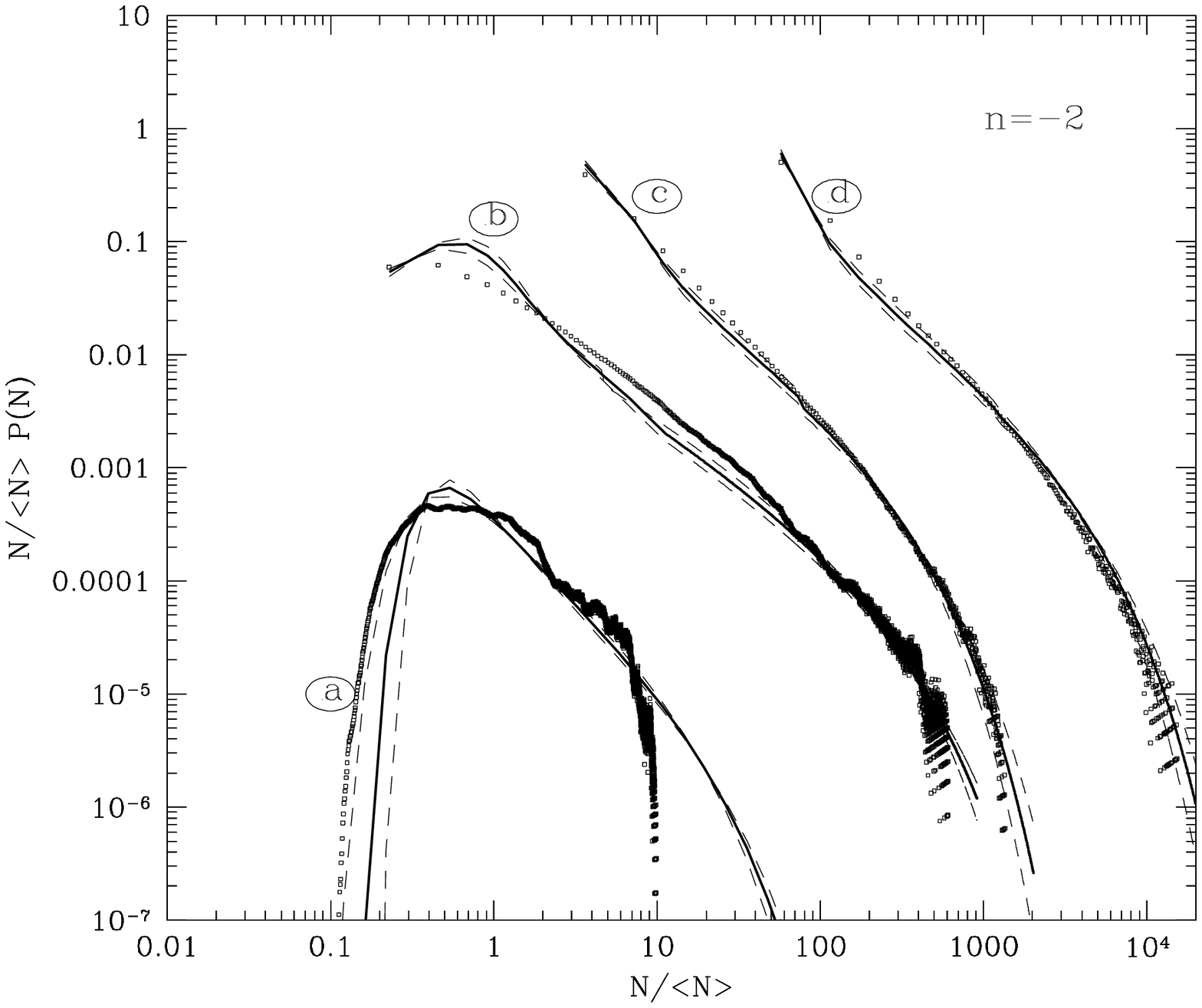,width=8.5cm}
}}
\caption[ ]{The measured CPDF as a function of $N/{\langle N \rangle}$ 
(small squares) for various values of $\Nb$ and $\sigma$ (see Table 2),
where $\Nb\equiv \sum_N N P_N$ is the average number of objects per cell. 
Each panel corresponds to a different value of $n_{\rm linear}$. The 
predicted CPDF from extended perturbation theory (eq.~[\ref{eq:ept}]) (solid
curve), is calculated assuming  that the 
function $n_{\rm eff}(\sigma)$ is given by equation 
(\ref{eq:fitber}) with the parameters displayed in 
Table~\ref{table:table1}. The two dashed curves bracketing the solid one 
correspond to the upper and the lower thin dashed curves 
in Figure \ref{fig:figure1}.}
\label{fig:figure2} 
\end{figure*} 
\noindent 
As described in the previous section, we find that the first few 
moments of the PDF
are well fitted by the predictions of EPT when the spectral index and the
variance are allowed to  be adjustable parameters. To see whether this
property holds at higher order, we compare in Figure \ref{fig:figure2}
the measured CPDF in our $N$-body sets (tiny squares) to what is
predicted by EPT (solid curves), for various values of $\sigma$ and
average number of object per cell
\begin{equation}
   \langle N \rangle \equiv \sum_{N=0}^{\infty} N P_N
\end{equation}
(see Table 2). The details of the computations of the PDF in the EPT
framework are given in detail in Appendix A. To obtain the CPDF, one
convolves the final result with a Poisson distribution
(see, e.g., Coles \& Jones 1991). Concerning the function
$n_{\rm eff}(\sigma)$, we take the fit (\ref{eq:fitber}) with the
parameters given in Table~\ref{table:table1}.  The dashed curves use
the functions $n_{\rm eff}(\sigma)$ corresponding to the thin
short-dashed curves in Figure \ref{fig:figure1}. To some extent, they
reflect the uncertainty in the value of $n_{\rm eff}$. However, there
are other significant sources of error, which we list here:
\begin{itemize} 
\item There is an uncertainty in the value of $\sigma$ used to 
compute the CPDF in the EPT framework, since we
use $\sigma$ as measured in the $N$-body experiments. 
For $n_{\rm linear} \geq -1$, the error in this estimate is expected 
to be rather small, less than or of order $30\%$.
For $n_{\rm linear}=-2$, finite volume effects strongly contaminate 
the measurements (CBH). 
In that case, we take the 
polynomial fit for $\sigma^2$ (as a function of scale) 
in logarithmic coordinates computed
by CBH, which is 
expected to give values of $\sigma^2$ correct to $\sim 30\%$. 
We do not take explicitly into account the uncertainty in the measurement of
$\sigma$ in what follows, although it is by definition a requirement to know
$\sigma$ accurately for EPT to give sensible predictions. Indeed, since 
the variance strongly influences the shape of the PDF, particularly its width,
it is obvious that two PDFs with different values of $\sigma^2$ 
cannot be expected to agree with each other (unless one considers a finite 
range  of values of $\rho$ for the comparison). 

\item There is an uncertainty in the measurement of the CPDF itself. 
In particular, the tails of the CPDF are subject to a variety of
spurious effects (discussed in detail in CBSI, CBSII and CBH):
(i) some memory of the
pattern used to set up the initial conditions is
likely to be preserved
during the simulation, particularly in under-dense regions.
In our case, particles are initially perturbed from a
mesh, inducing {\em grid effects}
which are expected to alter the measured CPDF at
small values of $N$, increasing
with $-n_{\rm linear}$ (CBH). These effects can in principle be
avoided by using a ``glass'' to generate initial
conditions\footnote{Of course, 
this procedure does not correct for the discrete nature of the simulated
data but keeps the white noise level of the particles as low as possible, without
introducing grid effects (see, e.g., Baugh et al. 1995).}
(White 1994). 
(ii) Because of the finite size of the simulation box, the large-$N$
tail of the CPDF is dominated by a few rare clusters, inducing
larger and larger (spurious) fluctuations in the CPDF as $N$ 
increases, followed by a sharp cut-off. 
\end{itemize}

\begin{table}
\caption[ ]{Values of $\sigma$ and $\Nb$ used for the curves in Figure 2.}
\label{table:table2}
\begin{tabular}{ccccccc}
\hline
 $n_{\rm linear}$  & curves     &  (a)  &  (b)  &  (c)  &  (d)  &  (e) \\ 
 $-2$ & $\sigma$   &  1.49 &  5.28 &  5.1  & 16.2  &      \\
                     & $\Nb$      & 1100 & 4.37 & 0.276 & 0.0174 &    \\ \hline
 $-1$ &  $\sigma$  & 0.57 & 0.56 & 3.72 & 3.36  & 17.94  \\
                     & $\Nb$   & 1100 & 69.3 & 4.37 & 0.276 & 0.0174 \\ \hline
 $0$  &  $\sigma$  & 0.38 & 3.67 & 3.46  & 22.9   &    \\
                     & $\Nb$      & 1100 & 4.37 & 0.276 & 0.0174 &    \\ \hline
 $1$  & $\sigma$   & 0.32 & 4.19 & 4.0   & 30.3   &    \\
                     & $\Nb$      & 1100 & 4.37 & 0.276 & 0.0174 &    \\ \hline
\end{tabular}
\end{table}

With this in mind, we see that the agreement between EPT and the
measurements is excellent, except for $n_{\rm linear}=-2$, a case we
discuss further below.  There are some very small discrepancies at the
largest value of $\sigma$ we analyzed for $n_{\rm linear}=0$ and 1
[right curves of top panels of Fig.~\ref{fig:figure2}, with label
(d)]. Otherwise, the solid curves almost exactly superpose on the
small squares.

For $n_{\rm linear}=-2$, there are some significant disagreements
between EPT and the measurements.  While the measured high-$N$ tail of
the CPDF is quite well described by EPT, there is disagreement in the
second curve from the bottom [label (b)], where the shape of the CPDF
predicted by EPT appears to be incorrect at small and moderate $N$.
For the lowest curve [label (a)], the agreement between EPT and
measurements is not as bad as it looks, given the uncertainties in the
function $n_{\rm eff}$ and the large errors in the measured CPDF
arising from finite volume effects.  This is actually not surprising
since the results are nearly in the weakly nonlinear regime where PT
is expected to apply.

To summarize our $n_{\rm linear}=-2$ analysis, EPT does not seem to
describe very well what is happening at small and moderate $N$, i.e.
in under-dense regions, but is still quite accurate for large values
of $N$, i.e. in over-dense regions. This result is not in
contradiction with the fact that we infer a single function $n_{\rm
eff}(\sigma)$.  Indeed, we measure it from low-order moments of the
CPDF, which are dominated by the large-$N$ tail of the CPDF,
increasingly so with the order $q$.

\subsection{Application to non power-law spectra}

%
%
\begin{figure}
\centerline{\psfig{figure=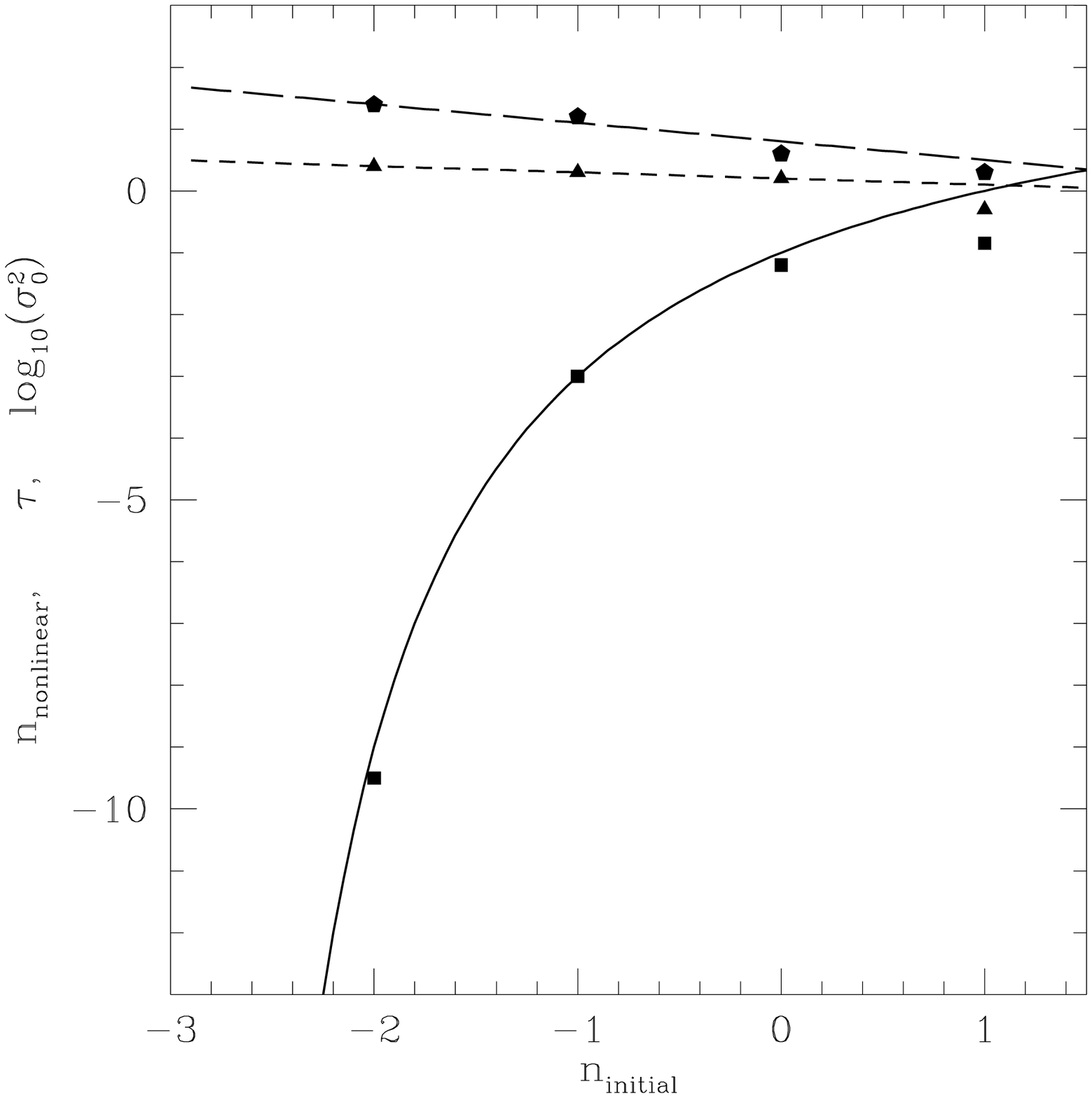,width=8.5cm}}
\caption[ ]{The $n_{\rm linear}$ dependence
of the parameters $n_{\rm nonlinear}$ (squares), $\tau$ (pentagons) and 
$\log_{10} \sigma^2_0$ (triangles) used
in the fit I [eq.~(\ref{eq:fitber})]. The symbols correspond to the 
values given in Table~2. The solid curve, long dashes and short dashes
correspond to the simple analytical forms (\ref{eq:fitnnl}), (\ref{eq:fittau})
and (\ref{eq:fits0}) respectively.}
\label{fig:figure3} 
\end{figure} 

\begin{figure}
\centerline{\psfig{figure=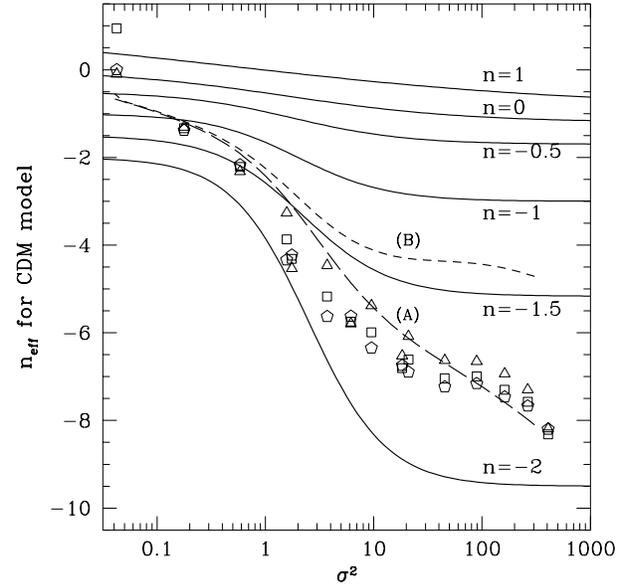,width=8.5cm}}
\caption[ ]{The effective index for the CDM model. The triangles,
squares and pentagons correspond to the measured functions $n_q(\sigma)$ for
$q=3,4$ and 5 respectively. The solid lines correspond to our 
phenomenological fit $n_{\rm eff}^{\rm I}$ for scale-free initial
conditions [eq.~(\ref{eq:fitber})] with
parameters given by Table~\ref{table:table1}, and by
eqs.~(\ref{eq:fitnnl}), (\ref{eq:fittau}) and (\ref{eq:fits0}) for $n_{\rm linear}=-0.5$
and $n_{\rm linear}=-1.5$. The numbers $n$ on the figure stand for $n_{\rm linear}$.
The long dashes correspond to the prediction for CDM using the prescription
(A) and the short dashes use the prescription (B) (see text).}
\label{fig:figure4} 
\end{figure}

The previous results apply to scale-free power spectra with particular
values of $n_{\rm linear}$. The first step to generalize them to
arbitrary (Gaussian) initial conditions is to extrapolate them to
other values of $n_{\rm linear}$. Here, we propose a series of simple
fits for the values of the parameters $n_{\rm nonlinear}$, $\sigma_0$
and $\tau$, of fit I as functions of $n_{\rm linear}$:
\begin{eqnarray}
   n_{\rm nonlinear}(n_{\rm linear})&=&\frac{3\ (n_{\rm linear}-1)}{3+n_{\rm linear}}
  \label{eq:fitnnl}\\
\tau(n_{\rm linear})&=&0.8-0.3\ n_{\rm linear}\label{eq:fittau}\\
\log_{10} \sigma^2_0(n_{\rm linear})&=&0.2-0.1\ n_{\rm linear}\label{eq:fits0}
\end{eqnarray}
As can been seen from Figure~\ref{fig:figure3}, these fits reproduce well
the values found for $n_{\rm linear}=-2$, $-1$ and $0$, but are less
accurate for $n_{\rm linear}=+1$.  In any event, the values of 
cosmological interest
are within the range $n_{\rm linear}\leq 0$.
Figure~\ref{fig:figure4} shows the quantities $n_{\rm eff}^{\rm I}$
obtained for different values of the index $n_{\rm linear}$
as functions of the variance (solid lines). [The interpolations
(\ref{eq:fitnnl}), (\ref{eq:fittau}) and (\ref{eq:fits0}) are used
only for $n_{\rm linear}=-0.5$ and $n_{\rm linear}=-1.5$.]

Note that Fry, Mellot \& Shandarin (1993) 
measured $S_3 \simeq 9/(3+n_{\rm linear})$
in the highly nonlinear regime, which implies $n_{\rm nonlinear}\simeq
(1.9 n_{\rm linear} -2.5)/(3+n_{\rm linear})$, a formula very similar
to equation~(\ref{eq:fitnnl}).  This equation diverges for $n_{\rm
linear}=-3$, implying that equation (\ref{eq:fitnnl}) is probably
valid only for $-3 < n_{\rm linear} \la 0$.

Now, we can generalize the above results to scale dependent
power-spectra.  This is actually not completely straightforward. For
scale-free initial conditions, a single parameter $n_{\rm linear}$ can
be used to entirely describe the shape of the initial power-spectrum.
For more complex initial conditions, such as CDM, PT predictions
require not only a knowledge of $-(n_{\rm linear}+3) =
d\log\sigma_{\rm linear}^2/d\log \ell$, but also of higher logarithmic
derivatives of the linear variance with respect to the filtering scale
\begin{equation}
  \gamma_p \equiv
\frac{d \log^p \sigma^2_{\rm linear}}{d \log^p \ell}, \quad p \geq 2.
\end{equation}
For example, equation (\ref{eq:S4per}) must now include a
term depending on $\gamma_2$ and more generally, the ratio $S_q$ involves
$\gamma_p$ terms for $p \leq q-2$.  It was, however, noticed by
Bernardeau (1994b) that for CDM like power spectra, such corrections
can be neglected for $S_q$, $q\leq 4$, i.e., one can take $\gamma_p
=0$ for $p \geq 2$, while still preserving reasonable accuracy.
This may not be true when higher order moments must be predicted
accurately.  Baugh et al. (1995) took into account
$\gamma_p$ for $p\leq 3$ to predict $S_q$ up to $q=10$.  But these are
tiny refinements and it has been shown that, in the weakly nonlinear
regime, the shape of the PDF is quite well reproduced when only the
local index $n_{\rm linear}$ is used (B94). There is no
reason why this should also be the case in the EPT framework, but we
can assume that it is so and check this hypothesis directly by measuring
the functions $n_q$ in simulations with scale-dependent initial
conditions.

%
%
Figure~\ref{fig:figure4} displays the functions $n_q(\sigma)$ obtained
from two CDM simulations with different box sizes (by Davis \&
Efstathiou 1988 and Frenk \etal 1990), using measurements of the
ratios $S_3$ (triangles), $S_4$ (squares) and $S_5$ (pentagons) by
CBSI. Again, as in Figure~\ref{fig:figure1}, the quantities $n_q$
superpose to define a unique function $n_{\rm eff}(\sigma)$, in
agreement with EPT.

Now it remains to be seen whether one can infer the function $n_{\rm
eff}(\sigma)$ from our results for scale-free power-spectra [using
eqs.~(\ref{eq:fitnnl}), (\ref{eq:fittau}) and (\ref{eq:fits0})]. The
problem is that the quantity $n_{\rm linear}$ now depends on scale. In
the weakly nonlinear regime, we have of course
\begin{equation}
   n_{\rm linear}(\ell) = -\gamma_1(\ell)-3, \quad \sigma^2 \ll 1.
   \label{eq:A}
\end{equation}
But in the nonlinear regime there is an ambiguity concerning the scale
at which the initial index should be chosen. 
We consider two simple options:
\begin{description}
\item[{\rm (A)}] With an Eulerian approach, one would generalize formula (\ref{eq:A})
to any value of $\sigma$.
\item[{\rm (B)}] Within a Lagrangian framework, one would rather
choose $n_{\rm linear}$ according to
\begin{equation}
   n_{\rm linear}(\ell)\equiv -\gamma_1(\ell_s)-3,
\end{equation}
where the scale $\ell_s$ is related to $\ell$ in the same way as in the
Hamilton et al. (1991) prescription for the transform function
of the variance, \ie
\begin{equation}
   \ell_s^3\approx \ell^3\ (1+\sigma^2).
\end{equation}
Of course, we have $\ell_s \simeq \ell$ when $\sigma^2 \ll 1$, in agreement with
equation~(\ref{eq:A}).
\end{description}
In Figure~\ref{fig:figure4}, we compare the prescriptions (A) (long
dashes) and (B) (short dashes) to our measurement of function $n_{\rm
eff}(\sigma)$.  The first prescription appears to be clearly better
than the second. This is somewhat surprising since prescription (B) is
expected to be more ``physical'' than (A). Further investigations are
required to understand the implications of this result.

\section{The Density PDF in the Highly Nonlinear Regime}
In the continuous limit, and in the highly nonlinear regime, one can
use PT results to compute some simple analytical properties of the PDF
in the EPT framework. Indeed, as we are going to see in \S~3.1, the
full shape and properties of the PDF in this regime can be simply
described by a few parameters, which are of course all fixed by the
value of $n_{\rm eff}$. One can thus check EPT predictions by
measuring, if possible, these parameters directly from the shape of
the PDF, and by comparing the results with what would be obtained with
the values of $n_{\rm eff}$ we measured in \S~2.1. This analysis,
which is the subject of \S~3.2, should confirm the results of \S~2.2.

\subsection{EPT results}

Here, extrapolating the PT calculations of B94 to the highly nonlinear
regime (i.e. assuming $S_q\equiv S_q^{\rm PT}$), 
we sketch simple analytical properties of the density PDF
in the EPT framework, when the variance is large. Details are given
in Appendix A. Let us just recall 
that in the continuous limit, the shape of the density PDF
is related to two functions $g(z)$ and $h(x)$, which are both
entirely determined from the expression of the generating function
of the $S_q$ parameters (BS). 
Their respective domains of validity are bounded by
\begin{equation}
 \rho_o=\sigma^2
\end{equation}
and
\begin{equation}
 \rho_u=a^{1/(1-\omega)}\sigma^{2\omega/(\omega-1)},
\end{equation}
where the parameters $a$ and $\omega$ are set by the value of
$n_{\rm eff}$ (see Table 3 for numerical values and Appendix A.1 for
analytical expressions of $a$ and $\omega$). There are two regimes:
\begin{eqnarray}
   P_{\rm PT}(\rho,\sigma,n_{\rm eff})&\simeq&
   \frac{1}{\rho_u} g\left( \frac{\rho}{\rho_u} \right), 
    \quad {\rm when}\ \ \rho \ll \rho_o,
   \label{eq:foncg}
\end{eqnarray}
and
\begin{eqnarray}
   P_{\rm PT}(\rho,\sigma,n_{\rm eff})&\simeq&
   \frac{1}{\rho_o^2} h\left( \frac{\rho}{\rho_o} \right), 
   \quad {\rm when}\ \ \rho \gg \rho_u.
   \label{eq:fonch}
\end{eqnarray}
These regimes actually overlap in the region $\rho_u\ll\rho\ll\rho_o$
when the variance is large. In this interval, functions $g$ and $h$
exhibit the same power law behavior. Note that the global shape
of $g(z)$ depends only on the value of $\omega$ (see BS, B94 and Appendix A).
It exhibits an exponential cut-off at small $z$.
The  function $h(x)$ is more complex and depends 
on the whole shape of the generating function of $S_q$.
It is expected to exhibit the following power-law behavior at small $x$ 
\begin{equation}
  h(x) \simeq a \frac{(1-\omega)}{\Gamma(\omega)} x^{\omega-2}, \quad x \ll 1,
  \label{eq:mxll1}
\end{equation}
and an exponential tail at large $x$
\begin{equation}
  h(x) \simeq 
   \frac{ 3 a_s}{4 \sqrt{\pi}} x^{-5/2} \exp( -|y_s| x), \quad x \gg 1.
 \label{eq:mxgg1}
\end{equation}
Again, the parameters $y_s$ and $a_s$ are entirely fixed by the value
of $n_{\rm eff}$ (see Table~\ref{table:table3} and Appendix A.1 for the
analytical expression of $y_s$).

\begin{figure*}
\centerline{
\hbox{
\psfig{figure=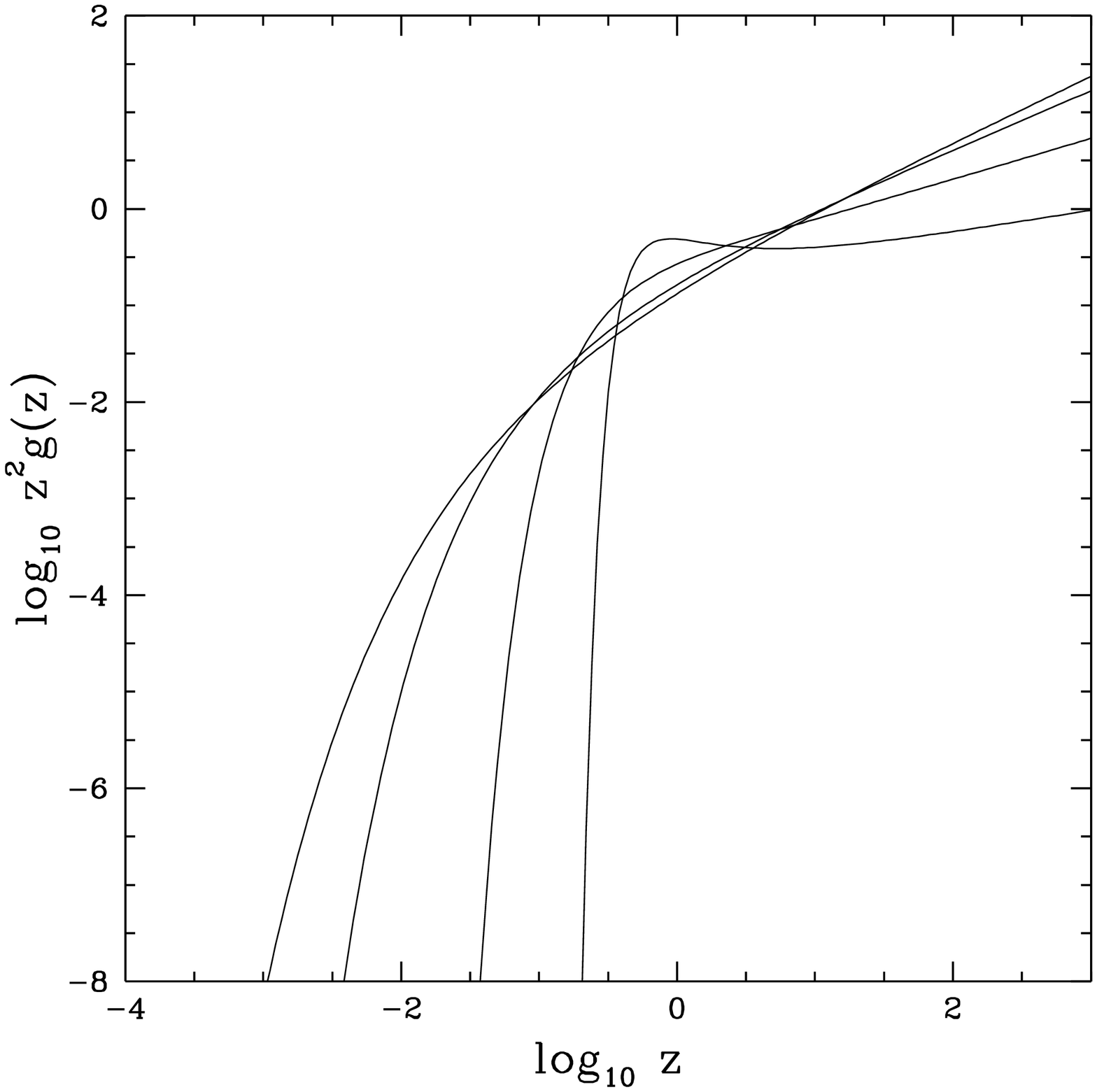,width=8cm}
\psfig{figure=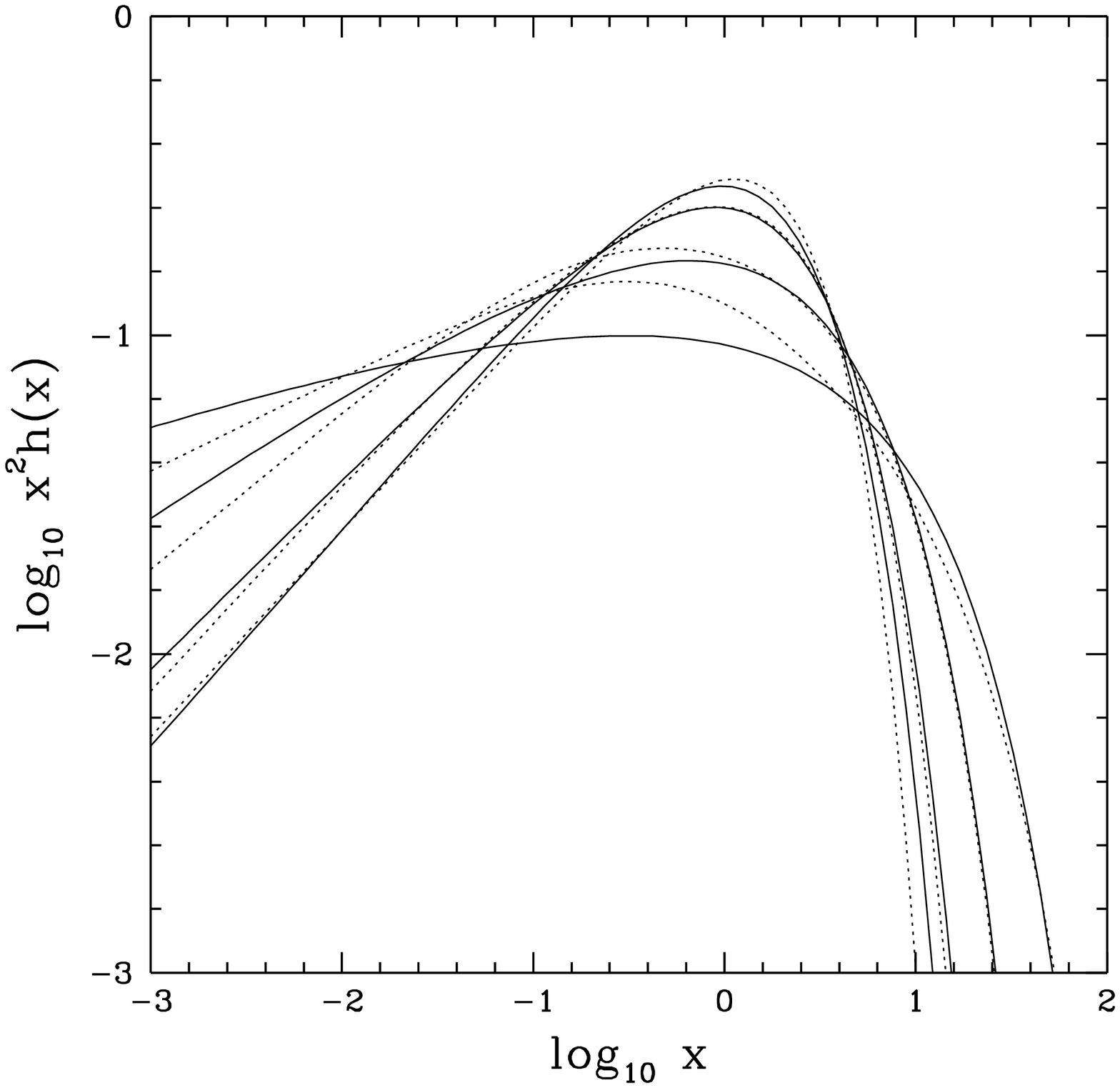,width=8.2cm}
}}
\caption[ ]{The functions $z^2g(z)$ (left panel) and $x^2h(x)$ (right panel),
in logarithmic coordinates, for various values of $n_{\rm eff}=-0.5$,
$-1$, $-3$ and $-9$ (solid curves). When $n_{\rm eff}$ decreases, the
exponential tail of the function $z^2g(z)$ at small $z$ becomes
sharper and the power-law at large $z$ less steep.
The maximum of $x^2h(x)$ increases with $n_{\rm eff}$. The dashed
curves are the ``preferred'' fits of the function $h$ 
measured in our $N$-body simulations, using equation (\ref{eq:hfit})
with $a$, $b$, $c$,  $\omega$ and $|y_s|$ as adjustable parameters. If
EPT applies, each dashed curve should, 
within measurement errors, superpose to the globally closest solid curve.}
\label{fig:fonction_gh} 
\end{figure*}

Figure \ref{fig:fonction_gh} shows the functions $z^2 g(z)$ and $x^2 h(x)$ 
[numerically obtained from equations~(\ref{eq:foncge}) and
(\ref{eq:fonche}) in Appendix A.1] 
in logarithmic coordinates
for various values of $n_{\rm eff}=-9$, $-3$, $-1$ and $-0.5$. 
These are approximately the values
of $n_{\rm eff}(\sigma)$ we measured in our scale-free simulations at
$\sigma=10$ for $n_{\rm linear}=-2$, $-1$, 0
and +1 respectively. 

\begin{table}
\caption[ ]{Parameters $\omega$, $a$, $a_s$ and $|y_s|$ 
for various values of the spectral index $n_{\rm eff}$.}
\label{table:table3}
\begin{tabular}{ccccc}
\hline
$n_{\rm eff}$   & $\omega$ & $a$ & $a_s$ & $1/|y_s|$ \\ \hline
-0.5  & 0.6667   & 2.044 & 18.94 & 1.7436  \\
-1    & 0.6000   & 2.104 & 6.677 & 2.4923  \\
-2    & 0.5000   & 2.121 & 2.809 & 3.9574  \\
-3    & 0.4286   & 2.082 & 1.842 & 5.4112  \\
-4    & 0.3750   & 2.025 & 1.413 & 6.8614  \\
-5    & 0.3333   & 1.966 & 1.169 & 8.3100  \\
-6    & 0.3000   & 1.908 & 1.012 & 9.7576  \\
-7    & 0.2727   & 1.855 & 0.900 & 11.205  \\
-9    & 0.2308   & 1.763 & 0.752 & 14.098  \\
-11   & 0.2000   & 1.689 & 0.657 & 16.991  \\ \hline
\end{tabular}
\end{table}
%
%
%

\subsection{Comparisons with $N$-body Results}
%

As for the CPDF in \S~2.2, we can check the validity of EPT by testing
directly the analytical forms of \S~3.1 in our $N$-body simulations. In
particular, we can try to measure the parameters fixing the shape of
the functions $g$ and $h$ (if they exist, of course), and compare the
results with what would be obtained from the values of $n_{\rm eff}$
we measured in \S~2.1. This analysis should confirm the results of
\S~2.2.  However, it is not easy to carry this out, for several reasons:
\begin{itemize}
\item The effective power index $n_{\rm eff}$ can vary with 
$\sigma$ even in the highly nonlinear regime, which
would render the comparisons with EPT quite difficult, since 
each value of the variance should be considered
independently. However this variation is, as discussed in CBH and in \S~2.1, 
quite weak, and $n_{\rm eff}$ can be taken to be constant to
a first order of approximation. 
\item Strictly speaking, the continuous limit is not reached 
in our $N$-body simulations. 
Discreteness effects can significantly contaminate the measurements, 
even in samples with $\sim 3\times 10^5$ objects like
our $N$-body sets. Fortunately one can correct for them 
(at least partially), as proposed by Bouchet, Schaeffer \& Davis 
(1991, hereafter BSD). 
\item As discussed in \S~2.2, other spurious effects, such 
as grid effects at small $N$ and more importantly finite volume effects 
at large $N$ can significantly affect the measurements.
\end{itemize}
Except for the first point above, these problems have been 
extensively discussed in BSD, Bouchet \& Hernquist (1992,
hereafter BH) and particularly in CBSII. We refer the
reader to these 
papers for details concerning the practical measurement
of the functions $g$ and $h$. 

Although we do not find it necessary to present the results here, we
find a remarkable qualitative agreement between the measurements and
the predictions (\ref{eq:foncg}) and (\ref{eq:fonch}) in the regime
$\sigma^2 \gg 1$, after the appropriate corrections. Aside from
the case $n_{\rm linear}=+1$, and to a lesser extent $n_{\rm linear}=0$, the
scale dependence of $n_{\rm eff}$ cannot really be neglected (see
Fig.~\ref{fig:figure1}).

We also tried to estimate by direct measurements (see, e.g., BSD, BH, CBSII)
the most important parameters of the functions $g$ and $h$, \ie 
$\omega$ and $|y_s|$. 
The first one, $\omega$, indeed completely determines the behavior of the PDF
around its maximum and in under-dense regions, and the second one, 
$|y_s|$,  determines the high-density tail of the PDF,
therefore influencing enormously the values of the moments of the PDF.  
The parameter $\omega$ is thus estimated by studying the 
small-$N$ behavior of the CPDF (through an estimator of the function
$g$, using the corrections for discreteness proposed by BSD), 
whereas $|y_s|$ is estimated by analyzing the large-$N$ 
behavior of the CPDF (through an estimator of the function $h$,
using the corresponding BSD corrections for discreteness). 
Table~\ref{table:param} summarizes the results of our measurements 
($\omega(g),1/|y_s(h)|$), 
compared with what is obtained
from equations (\ref{eq:omegan}) and (\ref{eq:ysn}) in Appendix A.1 taking for $n$ 
the value of $n_{\rm eff}$ we measure from
quantities $S_Q$ at $\sigma^2 \sim 100$ 
($\omega(n_{\rm eff}),1/|y_s(n_{\rm eff})|$). 
\begin{table*}
\caption[ ]{Comparison between EPT and measurements in our 
$N$-body simulations for $\omega$ and $|y_s|$.}
\label{table:param}
\begin{center}
\mbox{
\begin{tabular}{cccccc}
\hline
$n_{\rm linear}$  & $n_{\rm eff}$ & $\omega(g)$ &  $\omega(n_{\rm eff})$ & $1/|y_s(h)|$ & $1/|y_s(n_{\rm eff})|$ \\ \hline
$-2$               & $-9\pm^{2.18}_{2.75}$ & $0.3\pm^{0.05}_{0.03}$  & $0.23\pm^{0.05}_{0.04}$ & $18\pm^{2}_{6}$ & $14\pm^{4}_{3}$ \\
 $-1$               & $-3\pm^{0.67}_{0.79}$ & $0.5\pm0.05$            & $0.43\pm 0.04$          & $5\pm^{1}_{0.5}$ & $5.4\pm^{1.2}_{1}$ \\
 0                & $-1\pm^{0.32}_{0.36}$ & $0.65\pm^{0.15}_{0.1}$ & $0.6\pm 0.04$           & $2\pm^{0.5}_{0.2}$ & $2.5\pm{0.5}$ \\
 $+1$             & $-0.5\pm^{0.24}_{0.28}$ & $0.65\pm^{0.35}_{0.05}$ & $0.67\pm^{0.03}_{0.04}$ & $1\pm^{0.5}_{0}$   & $1.74\pm 0.4$ \\
\hline
\end{tabular}
}
\end{center}
\end{table*}

The agreement between both estimates of $\omega$ and $|y_s|$ 
is quite good, as expected if EPT applies. 
There is a small problem in the case $n_{\rm linear}=-2$, 
where the measured $\omega$ seems larger than the one
given by EPT (but the error bars on each estimate overlap). 
The same effect exists for $n_{\rm linear}=-1$, 
but it is less significant.

We also tried to find a best fit for the function $h$, using the 
parameterization proposed by BSD, \ie
\begin{equation}
  h_{\rm fit}(x)=a \frac{(1-\omega)}{\Gamma(\omega)} 
x^{\omega-2}(1+bx)^{-c} \exp(-|y_s| x),
  \label{eq:hfit}
\end{equation}
with $a$, $b$ and $c$ being {\em adjustable} parameters\footnote{Note 
that, in principle, in equation (\ref{eq:hfit}), 
the parameters $a$, $b$ and $c$ are entirely determined
in the EPT framework. The number $a$ is indeed given by equation (\ref{eq:vala})
in Appendix A.1 and from equation~(\ref{eq:mxgg1}), we get
\begin{equation}
  c=\omega+1/2,
\end{equation}
\begin{equation}
  b=\left[ \frac{ 4\sqrt{\pi} a(1-\omega)}
  {3 a_s \Gamma(\omega)} \right]^{1/(\omega+1/2)}.
\end{equation}
However, with such values of $a$, $b$ and $c$, the simple analytical form
(\ref{eq:hfit}) approximates  the exact EPT result satisfactorily only for 
$n_{\rm eff} > -2$. At the degree of approximation we consider here, it
is thus better to keep $a$, $b$ and $c$ as free parameters.}, and with 
our ``preferred'' values of $\omega$ and $|y_s|$
obtained from the direct measurement of the functions $g$ 
and $h$. The resulting parameters are listed
in Table~\ref{table:fitpref}. We obtained the dotted curves 
in the right panel of Figure~\ref{fig:fonction_gh}, which compare
well with the predictions from EPT, with our ``preferred'' 
measured value of $n_{\rm eff}$ at $\sigma^2 \sim 100$,
particularly for the large-$x$ exponential tail. 
The agreement with EPT is not that good at small $x$
for $n_{\rm eff}=-9$ ($n_{\rm linear}=-2$) and is marginal
for $n_{\rm eff}=-3$ ($n_{\rm linear}=-1$). 
Of course, all these measurements are quite delicate and 
the uncertainties are large. However, the disagreement
for $n_{\rm eff}=-9$ is quite consistent with the conclusions of
\S~2.2. 

Thus, the results of this section corroborate those of \S~2.2,
as expected.
\begin{table}
\caption[ ]{Our preferred values of $a$, $b$, $c$, $\omega$ 
and $|y_s|$ in equation (\ref{eq:hfit}) as we
measure functions $g$ and $h$.}
\label{table:fitpref}
\begin{tabular}{cccccc}
\hline
$n_{\rm linear}$ & $a$ & $b$ & $c$ & $\omega$ & $1/|y_s|$ \\ \hline
  $-2$             & 1.273 & 1.455 & 0.9 & 0.3  & 18        \\
  $-1$             & 2.068 & 4.3   & 0.6 & 0.5  & 5         \\
  0              & 2.7   & 8     & 0.225 & 0.65 & 2       \\
  $+1$             & 1.937 & 0.486 & $-$1.35 & 0.65 & 1       \\
\hline
\end{tabular}
\end{table}
%
%

%
\section{Discussion}
%
In this paper, we have studied the probability distribution function (PDF) of
the density field smoothed with top hat windows in $N$-body simulations 
with scale-free initial conditions and 
linear spectral indices $n_{\rm linear}=-2$, $-1$, 0 and $+1$. 
We attempted to determine if the 
following {\em Extended Perturbation Theory} (EPT) applies.  
Such an ansatz consists of assuming
that the PDF is given by $P(\rho)=P_{\rm PT}(\rho,\sigma,n_{\rm eff})$, 
where the function
$P_{\rm PT}(\rho,\sigma,n_{\rm linear})$ is the prediction of {\em Perturbation
Theory} (PT, e.g., B94), 
which is valid in the
weakly nonlinear regime $\sigma\simeq \sigma_{\rm linear} \ll 1$. 
The difference with PT is that
the {\em effective} spectral index $n_{\rm eff}$ is assumed to be 
an adjustable parameter of the model and
$\sigma^2$ is the {\em nonlinear} variance\footnote{Thus, the
PDF obtained from EPT has exactly the same variance as the
data, by definition. This is obviously a necessary condition for agreement between theory
and measurements (but not a sufficient one, in general, except
of course in the Gaussian limit). This is why we think it does not
make much sense to try extending PT results to the nonlinear regime
by identifying the variance of the theoretical PDF with the value given by
linear theory.}.
The measurement of low-order moments of the PDF and of the PDF itself 
in our $N$-body experiments shows
a remarkable agreement with EPT, except for $n_{\rm linear}=-2$ in 
some regimes. (This is however a limiting
case, subject to many numerical uncertainties, particularly finite 
volume effects, see e.g. CBH). 

As can been seen from Figure~\ref{fig:figure1}, the effective spectral
index, equal of course to $n_{\rm linear}$ in the weakly nonlinear
regime, decreases with $\sigma$ to approximately reach another plateau
at large $\sigma$.  If this second plateau were completely flat, our
measurements would agree with the predictions of stable clustering.  This
does not appear to be exactly the case, except perhaps for $n_{\rm
linear}=-2$, although the deviation from stable clustering 
is quite weak.  We do not discuss this problem further here, since it
was analyzed extensively in CBH.  The difference between the second
and the first plateau increases with $-n_{\rm linear}$. For $\sigma^2
\sim 100$, we have for example $n_{\rm eff} \simeq -9$, $-3$, $-1$ and
$-0.5$ respectively for $n_{\rm linear}=-2$, $-1$, 0 and +1.

It is important to notice that the quantity $-(3+n_{\rm eff})$ is 
{\em different} from the logarithmic
derivative of $\sigma^2$ with respect to scale, except of course 
in the weakly nonlinear regime. Indeed,
as measured in $N$-body simulations by Efstathiou et al. (1988), CBH,
and Jain (1996), the variance follows quite well
the stable clustering predictions in the framework of the self-similar solution
(see, e.g., Davis \& Peebles 1977; Peebles 1980), 
in the regime $\sigma^2 \ga 100$.
In particular, we get
\begin{equation}
 -(3+n_{\rm \sigma})\equiv\frac{d\log\sigma^2}{d\log \ell} 
\simeq-\frac{3(3+n_{\rm linear})}{5+n_{\rm linear}},\quad \sigma^2 \ga 100,
\end{equation}
i.e., $n_{\rm \sigma}=-2$, $-1.5$, $-1.2$, $-1$ respectively for 
$n_{\rm linear}=-2$, $-1$, 0 and $+1$. These values are quite
different from the $n_{\rm eff}$ we measure.

We have tried to apply EPT to a scale dependent power-spectrum, such
as CDM.  The results given in Figure~\ref{fig:figure4} are
encouraging. However, further investigations are needed to address the
case of non power-law spectra in greater detail.

Szapudi, Meiksin \& Nichol (1996) have recently measured the functions
$n_q$ in the Edinburgh/Durham Southern Galaxy catalogue (EDSGC). In
this angular survey, the functions $n_q$ superpose well, given the
uncertainties in the measurements, although there is a systematic
increase in the scatter of values of $n_q$ on small scales.  Szapudi
et al. find, as we do, that the function $n_{\rm eff}$ exhibits two
plateaus, one on small scales where $-7 \la n_{\rm eff} \la -4$ and
another on large scales where $-2.5 \la n_{\rm eff} \la -1$, in good
qualitative agreement with the results of our analysis and with what
would be expected in a CDM-like universe.  Of course, such a
comparison is limited by the fact that there are non trivial
projection effects of the galaxy distribution along the line of sight
in angular surveys. Moreover, one has to keep in mind that light does
not necessarily trace mass.

If EPT is indeed valid, a knowledge of the variance $\sigma$ and of
the spectral index $n_{\rm eff}$ is sufficient to compute the PDF. We
have provided analytical fits to the function $n_{\rm eff}(\sigma)$.
The nonlinear variance $\sigma^2$ can itself be calculated with other
nonlinear ansatz, such as the fitting formula found by Hamilton et al.
(1991) (improved later by Peacock \& Dodds 1994; Jain, Mo \& White
1995; Baugh \& Gazta\~naga 1996), provided one generalizes it to the
smoothed density field.

Our analysis has been done for flat universes with density parameter
$\Omega=1$. PT predictions for the ratios $S_q$ are expected to be
only weakly dependent on the value of $\Omega$ or on the cosmological
constant $\Lambda$ (e.g., Martel \& Freudling 1991, Bouchet \etal
1992, 1995, Bernardeau 1992, 1994b, Hivon \etal 1995). But this does not
really guaranty the existence of a universal function $n_{\rm
eff}(\sigma)$ in the case $\Omega\neq 1$ and/or $\Lambda \neq 0$. More
investigations have to be undertaken in this regime.

Of course, it also remains to explain why the hierarchy of cumulants
of the PDF looks similar in the weakly nonlinear and the highly
nonlinear regimes. One way to address this question would be to study
the transition between the two regimes, by computing higher order
corrections in equation (\ref{eq:Sqpt}) (e.g., Scoccimarro \& Frieman
1996; Scoccimarro 1996). Alternatively, as in the early attempts 
of Fry (1984) and
Hamilton (1988), one could study the BBGKY equations (see, e.g.,
Peebles 1980) to try to find clues for the determination of the $S_q$
parameters in the highly nonlinear regime. 

\section{Acknowledgments}
We thank E. Gazta\~naga for useful comments.
F. Bernardeau would like to thank IAP, where a large part of the work has 
been completed, for its warm hospitality.
This work was supported in part by the Pittsburgh Supercomputing
Center, the National Center for Supercomputing Applications
(Illinois), the San Diego Supercomputing Center and by an allocation
from the scientific counsel of IDRIS, Palaiseau.
L. Hernquist acknowledges support from 
NASA Theory Grant NAGW-2422, and the NSF under Grant ASC
93-18185 and the Presidential Faculty Fellows Program.
%
\section*{APPENDIX}
%
\section*{Practical calculation of the PDF in the 
framework of extended perturbation theory} 
%
In this appendix, we first study, in the framework of EPT, the derivation
of the cosmic density PDF, and its properties when the variance is large.
All the results are discussed in term of the index $n$.
For PT, $n$ is simply the initial 
spectral index, $n\equiv n_{\rm linear}$, and,
more generally, in EPT, we have $n \equiv n_{\rm eff}$.
In the following we will suppose that $n < 0$. 

\subsection*{A.1 Analytic predictions}
%
In general the density PDF is given by an inverse Laplace transform,
\begin{equation}
   P(\rho)\ \d\rho=
   \d\rho\int_{-\ii\infty}^{+\ii\infty}
   {\d y\over 2\pi\ii\sigma^2}
   \ \exp\left[-{\phi(y)\over \sigma^2}+{\rho\ y\over\sigma^2}
   \right] ,\label{eq:LapInv}
\end{equation}
where $\phi(y)$ is the generating function of the $S_q$ parameters
[eq.~(\ref{eq:sqdef})]:
\begin{equation}
  \phy(y) \equiv\sum_{q=1}^{\infty} (-1)^{q-1} \frac{S_q}{q!} y^q 
   \label{eq:sigser}
\end{equation}
($S_1\equiv S_2 \equiv 1$).

In the PT framework and for scale-free initial conditions, 
the function $\phy$ is determined by the following set of 
implicit equations\footnote{These equations have been obtained assuming
that the spherical collapse solution is well approximated by $\G=
(1-3\delta_i/2)^{-3/2}$, where $\delta_i$ is the linear density contrast
and $\G-1$ is the present density contrast.} (see B94):
\begin{eqnarray}
  y&=&-\frac{3}{8}(1+\G)^{(n-4)/3}\left[ (1+\G)^{2/3} -1 \right] \times \nonumber\\
   & & \left[ (n+3) (1+\G)^{2/3}+ 1-n  \right],
  \label{eq:yimp}
\end{eqnarray}
\begin{eqnarray}
 \phy&=&\frac{3}{8}(1+\G)^{(n-4)/3} \left[ (1+\G)^{2/3} -1 \right]
\times \nonumber\\
 & & (1+\G)\left[ -n (1+\G)^{2/3} +(n-4)\right].
  \label{eq:phyimp}
\end{eqnarray}
The integral (\ref{eq:LapInv}) is not easy to estimate
analytically. In general, one must calculate it numerically (see section A.2).
The important properties of the PDF, however, depend only on the behavior 
of $\phy(y)$ at large $y$ and near its
critical point $y_s$ (BS). The absolute value of $y_s$ is  
the radius of absolute convergence of the 
series (\ref{eq:sigser}).
From equations (\ref{eq:yimp}) and (\ref{eq:phyimp}), one can deduce
that when $y \gg 1$
\begin{equation}
  \phy(y)\simeq a y^{1-\omega},
\end{equation}
with
\begin{equation}
   a=\frac{4-n}{1-n} \left[ \frac{8}{3(1-n)} \right]^{3/(n-4)},
   \label{eq:vala}
\end{equation}
and
\begin{equation}
   \omega=\frac{3}{4-n}.
   \label{eq:omegan}
\end{equation}
On the other hand, the behavior of $\phy$ around $y_s$ can be written
\begin{equation}
  \phy(y)-\phy_s \simeq r_s (y-y_s) + a_s (y-y_s)^{3/2}.
\end{equation}
The calculation of $y_s$ yields
\begin{equation}
   y_s = -\frac{3}{8} A^{-n/2} (A-1) \left[(n-1)A-(n+3) \right],
   \label{eq:ysn}
\end{equation}
where
\begin{eqnarray}
   A &= &  \frac{\left[ (n^2-n-2)^2-(n-1)(n-4)n(n+3) \right]^{1/2}}
       {(n-1)(n-4)}\nonumber\\
  & &+\frac{ n^2 - n -2 }{(n-1)(n-4)}.
\end{eqnarray}
The value of $a_s$ is hard to obtain analytically and 
we compute it numerically for various values of $n$ 
(see Table~\ref{table:table3}). A knowledge
of $\phy_s$ and $r_s$ is not relevant here. 

We can now define, following the formalism of BS, 
the typical density of an under-dense region as
\begin{equation}
 \rho_u=a^{1/(1-\omega)}\sigma^{2\omega/(\omega-1)},
\end{equation}
and the typical density of an over-dense region as
\begin{equation}
 \rho_o=\sigma^2.
\end{equation}

For regions with small enough density, the PDF can
be written as
\begin{equation}
   P_{\rm PT}(\rho,\sigma,n)\simeq\frac{1}{\rho_u} 
g\left( \frac{\rho}{\rho_u} \right), \quad \rho \ll \rho_o.
\end{equation}
The shape of the function $g(z)$ depends only on the value of the 
parameter $\omega$ (BS) through
the following integral
\begin{eqnarray}
   g(z)&=& \displaystyle \frac{1}{\pi} \int_0^{\infty} du \exp[ -zu + u^{1-\omega} 
   \cos \pi \omega ] \times \nonumber\\
   & & \sin[ u^{1-\omega} \sin\pi \omega].
   \label{eq:foncge}
\end{eqnarray}
It has a exponential tail at small $z$ 
and a power-law behavior in $z^{\omega-2}$
at large $z$ (see B94 or BS for details). 

%
%

For regions with large enough density, we have
\begin{equation}
   P_{\rm PT}(\rho,\sigma,n)\simeq\frac{1}{\rho_o^2} 
h\left( \frac{\rho}{\rho_o} \right), \quad \rho \gg \rho_u,
\end{equation}
where the function $h(x)$ is related to function $\phy(y)$ 
through the following transform (BS)
\begin{equation}
  h(x)=-\frac{1}{2\pi} \int_{-\infty}^{+\infty} du\ \phy(iu)\ e^{iux}.
  \label{eq:fonche}
\end{equation}
It exhibits a power-law behavior (\ref{eq:mxll1}) at small $x$ 
(this regimes overlaps the large $z$ regime of function $g$)
and an exponential tail at large $x$ given by equation (\ref{eq:mxgg1}).
Note that the parameters $S_q$ are simply the moments of the function $h$, 
\begin{equation}
  S_q = \int_0^{\infty} x^q h(x) dx,
  \label{eq:sqh}
\end{equation}
and we have the following normalizations
\begin{equation}
  \int_0^{\infty} xh(x)dx=\int_0^{\infty} x^2 h(x) dx=1.
  \label{eq:norm}
\end{equation}

\subsection*{A.2 Numerical Integration}

\begin{figure}
\centerline{\psfig{figure=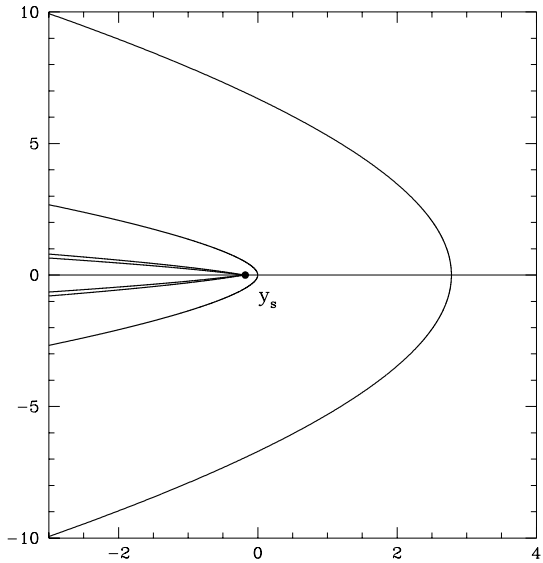,width=8.5cm}}
\caption[ ]{Example of paths in the complex plane for $y$. The parameters
correspond to $n=-3$ and $\sigma=1$. The paths cannot cross the real 
axis for $y<y_s$. When the density is small, the paths are on the
right side. When the density is large, they are on the left side and  
cross the real axis at $y=y_s$.}
\label{fig:ypath} 
\end{figure} 
The starting point of the numerical calculation of the density PDF is 
equation (\ref{eq:LapInv}). Since the intervening 
quantities are all known as functions
of $\G$, this latter is our natural variable of integration.
The expression for the density PDF thus reads,
\begin{eqnarray}
P(\rho)\d\rho & = & \displaystyle \d\rho\int_{-\ii\infty}^{+\ii\infty}
{\d \G\over 2\pi\ii\sigma^2}\ {\d y(\G)\over\d\G} \times \nonumber \\
& &\displaystyle \exp\left[-{\phi(y)\over \sigma^2}+{\rho\ y\over\sigma^2}
\right].\label{eq:LapInv2}
\end{eqnarray}
The difficult part of the integration is to choose the proper path in 
the complex 
plane. The original path for $y$ runs from $-\ii\infty$ to $+\ii\infty$
along the imaginary axis. But as the functions $\tau(y)$ or $\phi(y)$
are not analytic in $y$, there is at least one singularity on the
real axis for $y=y_s\ (<0)$, and it is impossible to move the path 
to the left of this singularity.  

Actually, the starting point of the path for the numerical integration
is the position of the saddle point, $y_{\rm saddle}$
defined by 
\begin{equation}
0\equiv\rho-\left.{\d\phi(y)\over\d y}\right\vert_{y=y_{\rm saddle}}.
\end{equation}
If there is no solution, that is when $\rho>1+\G_s$ for $y(\G_s)=y_s$,
the starting point is the singular point $y_s$.
Then, the path is chosen in such a way
that $\phi(y)+\rho y$ is kept real to avoid unnecessary oscillation
of the integrand.
Using this prescription, followed with an adaptive step, the paths that
are found are similar to the ones shown in Figure \ref{fig:ypath}.

%
%
%
%

\end{document}